\documentclass[a4paper]{aa}  
\usepackage{graphicx}
\usepackage{natbib}
\bibpunct{(}{)}{;}{a}{}{,}
\begin{document}
\title{Star formation efficiency in galaxy clusters}
\author{T. F. Lagan\'{a} \inst{1}
\and G. B. Lima Neto\inst{1} 
\and F. Andrade-Santos\inst{1}
\and E. S. Cypriano\inst{2}}

\institute{Universidade de S\~{a}o Paulo, Instituto de Astronomia, Geof\'{i}sica e Ci\^{e}ncias Atmosf\'{e}ricas, Departamento de Astronomia, Rua do Mat\~{a}o 1226, Cidade Universit\'{a}ria,
05508-090, S\~{a}o Paulo, SP, Brazil.
\and Department of Physics $\&$ Astronomy, University College London, London, WC1E 6BT, England.}

\date{Received  29 November 2007 / Accepted 5 April 2008}

\abstract
{The luminous material in clusters of galaxies exists in two forms: the visible galaxies and the X-ray emitting intra-cluster medium.
The hot intra-cluster gas is the major observed baryonic component of clusters, about six times more massive than the stellar component. The mass contained within visible galaxies is approximately 3$\%$ of the dynamical mass.}
{Our aim was to analyze both baryonic components, combining X-ray and optical data of a sample of five galaxy clusters (Abell 496, 1689, 2050, 2631 and 2667), within the redshift range 0.03 $ < {\it z} < $ 0.3. We determined the contribution of stars in galaxies and the intracluster medium to the total baryon budget.}
{We used public XMM-Newton data to determine the gas mass and to obtain the X-ray substructures. Using the optical counterparts from SDSS or CFHT we determined the stellar contribution.}
{We examine the relative contribution of galaxies, intra-cluster light and intra-cluster medium to baryon budget in clusters through the stellar-to-gas mass ratio, estimated with  recent data. We find that the stellar-to-gas mass ratio within $r_{500}$ (the radius within which the mean cluster density exceeds the critical density by a factor of 500), is anti-correlated with the ICM temperature, which range from 24$\%$ to 6$\%$ while the temperature ranges from 4.0 to 8.3 keV. This indicates that less massive cold clusters are more prolific star forming environments than massive hot clusters.}
{}

\keywords{galaxies: clusters: general-X-ray: galaxies: cluster-galaxies: luminosity function, mass function}

\maketitle 

\section{Introduction}
\label{intro}

Galaxy clusters occupy an unique position in the hierarchical scenario of structure formation as they are the largest bound and relaxed structures to form in the Universe \citep[e.g.,][]{tozzi07}. 
An important step in understanding galaxy clusters is to take into account all of their components. \citet{fukugita98} presented an estimate of the total budget of baryons in all states: most of the baryons are still in the form of ionized gas and the stars and remnants represent only 17$\%$ of the baryons. In a similar work \citet{ettori03} presented the cluster ``baryonic pie'' of which approximately 70$\%$ is composed of the hot intra-cluster medium and almost 13$\%$ composed of stars. 

The ratio of gas to stellar mass is particularly interesting because it is related to the star formation efficiency of clusters.  Studying the Hydra cluster, \citet{David90} computed the ratio of gas to stellar mass and it was found that the gas mass is almost 4 times that of stars. Using {\it Einstein} results from the literature, they extended this analysis to systems ranging from poor groups to rich clusters and a correlation between $M_{\rm gas}/M_{\rm stars}$ and the gas temperature was suggested.

\citet{LMS} analyzed a sample of 13 clusters, finding that the gas-to-stellar mass ratio increases from 5.9 to 10.4 from low- to high-mass clusters and their best-fit correlation for $M_{500}-L_{500}$ differs from $L_{500} \propto M_{500}$ by $3\sigma$, indicating a mass-to-light ratio increasing with mass. These two results strongly suggest a decrease of star formation efficiency in more massive environments. Recently, \citet{Gonzales07}, who  also used gas masses from the literature, estimated the contribution of stars in galaxies, intracluster stars and gas to the baryon budget. These authors, who took into account the intracluster light (ICL) to the baryon budget, confirming the previous trend of increasing $M_{\rm star}/M_{\rm gas}$ with decreasing temperature found by \citet{David90,LMS}. \citet{Roussel}, found that a typical stellar contribution to the baryonic mass is between 5$\%$ and 20$\%$, inside the virial radius, even though they claim that the stellar-to-gas mass ratio is roughly independent of temperature.

Although the baryon budget has been addressed in many studies 
\citep{David90,fukugita98,ettori03,Gonzales07} there are good reasons to revisit it. The most 
important reason is the stellar content contribution to the baryon fraction. Second, cosmological 
simulations can precisely predict the baryon fraction within $r_{500}$ (see Sect.~\ref{massdet}), 
which is approximately the radius in which the X-ray observations are reliable determined. 

To readdress this question, we choose five XMM-Newton public data clusters with SDSS or CFHT ({\it Canada-France-Hawaii Telescope}) counterparts to study the gaseous and the stellar baryon budgets. 

X-ray studies of galaxy clusters are particularly relevant to this issue as they provide the determination of the gas mass. An important task is to describe well the surface brightness profile, since the gas mass determination relies on it.

The baryonic stellar component can be estimated by integrating the luminosity function to obtain the total luminosity and then adding the intra-cluster light (ICL) contribution. Using an appropriate stellar mass-to-light ratio we have converted the total luminosity into stellar masses. 
 
The paper is organized as follows: Observations and data treatment are described in Sect.~\ref{obs}; we present the gas density distribution in Sect.~\ref{gasdist}; the temperature spectral analysis appears in Sect.~\ref{tempdist}; the dynamical analysis based on temperature and substructure maps is presented in Sect.~\ref{dynana}; the stellar baryonic determination is presented in Sect.~\ref{stelcont}; the mass determination is described in Sect.~\ref{massdet}. Finally we present our discussion in Sect.~\ref{discus} and we conclude in Sect.~\ref{conc}.

All distance-dependent quantities are derived assuming the Hubble constant
$H_{0}=70$~km~s$^{-1}$~Mpc$^{-1}$, $\Omega_{M}=0.3$ and, $\Omega_{\Lambda}$= 0.7.
Magnitudes are given in the AB system unless otherwise stated. All errors are
relative to a $68\%$ confidence level.

\section{Observations and data treatment}\label{obs}

\subsection{Sample}

The objects in our sample were drawn from a set of Abell clusters with both
XMM-Newton public archive and SDSS optical counterparts. We have also chosen
only clusters with redshifts in the range $0.03 < z < 0.3$ so that the cluster
image could be well resolved and were smaller than the detector field of view.
We thus had 6 clusters in this selection. Then we imposed that clusters have
exposure times of at least 10~ks after data reduction and filtering
(pipeline), and flare cleaning (see Sect.~\ref{RXana}). We ended up with 3
clusters: 1689, A2050 and A2631. We have not used either the temperature or
the object morphology as a criterion for cluster selection.

To enlarge this sample, we added the cluster A496 that was observed with the {\it Canada-France-Hawaii Telescope} (CFHT) and for A2667 we used total luminosity from the literature \citep{Covone06LK}. Properties of the whole sample are presented in Table~\ref{generalinfo}.

\subsection{X-rays}\label{RXana}

We used data from all EPIC cameras (MOS1, MOS2 and PN). For data reduction we used the XMM-Newton Science Analysis System (SAS) v6.0.4 and calibration database with all updates available prior to February, 2006.
The initial data screening was applied using recommended sets of event patterns, 0-12 and 0-4 for the MOS and PN cameras, respectively.

\begin{table*}[t!]
\centering
\caption{General cluster properties. Column~(1): cluster name; Col.~(2): right ascension; Col.~(3):declination; Col.~(4): redshift; Col.~(5): luminosity distance; Col.~(6): hydrogen column density \citep{DL}; Col.~(7): X-ray emission mean temperature derived in this work; see Sect.~\ref{tempmod}; Col.~(7): $r_{500}$ values were derived in this work.}
\begin{tabular}{cccccccc}
\hline\hline
 Cluster & R.A & DEC & ${\it z}$ & $d_{L}$ &nH & $<kT>$ & $r_{500}$ \\
         & (J2000) & (J2000)& & ($h_{\rm 70}^{-1}$ Mpc) & $(10^{20} cm^{-2})$ &  (keV) & ($h_{\rm 70}^{-1}$ kpc)\\
\hline\noalign{\smallskip}
A496  & 04 33 37.1 & -13 14 46 & 0.033  & 144.902 & 4.45 &  3.96 $\pm$ 0.023 & 1480 \\  
A1689 & 13 11 34.2 & -01 21 56 & 0.1823 & 888.870 & 1.84 &  8.34 $\pm$ 0.32 & 2172\\
A2050 & 15 16 21.6 & +00 05 59 & 0.1183 & 551.167 & 4.67 &  5.85 $\pm$ 0.11 & 1785   \\
A2631 & 23 37 39.7 & +00 17 37 & 0.273  & 1393.23 & 3.82 &  6.89 $\pm$ 0.34  & 2153    \\
A2667 & 23 51 47.1 & -26 00 18 & 0.23   & 1146.59 & 1.64 &  6.56 $\pm$ 0.13  & 1976\\
\hline	  
\end{tabular}
\label{generalinfo}
\end{table*}

The light curves are not constant and large variations in intensity are visible. These variations,  which we call flares, are caused by soft energy protons produced by solar activity. It is better to discard flare periods and reduce the effective exposure time to improve the signal-to-noise ratio. The light curves in the energy range of [1-10] keV were filtered to reject periods of high background. The cleaned light curves exhibited stable mean count rates and exposure times are given in Table \ref{obsinfo}. We considered events inside the field-of-view (FOV) and we excluded all bad pixels. 

In periods free of background flaring, the background is dominated by X-rays at low energies and particles at high energies. The X-ray component includes a significant contribution from Galactic emission which varies with position on the sky. Since cluster emission decreases with distance from the center, the background component becomes more important towards the outskirts. The background was taken into account by extracting MOS1, MOS2 and PN spectra from the publicly available EPIC blank sky templates described by \citet{Lumb02}.
 
\begin{figure*}[t!]
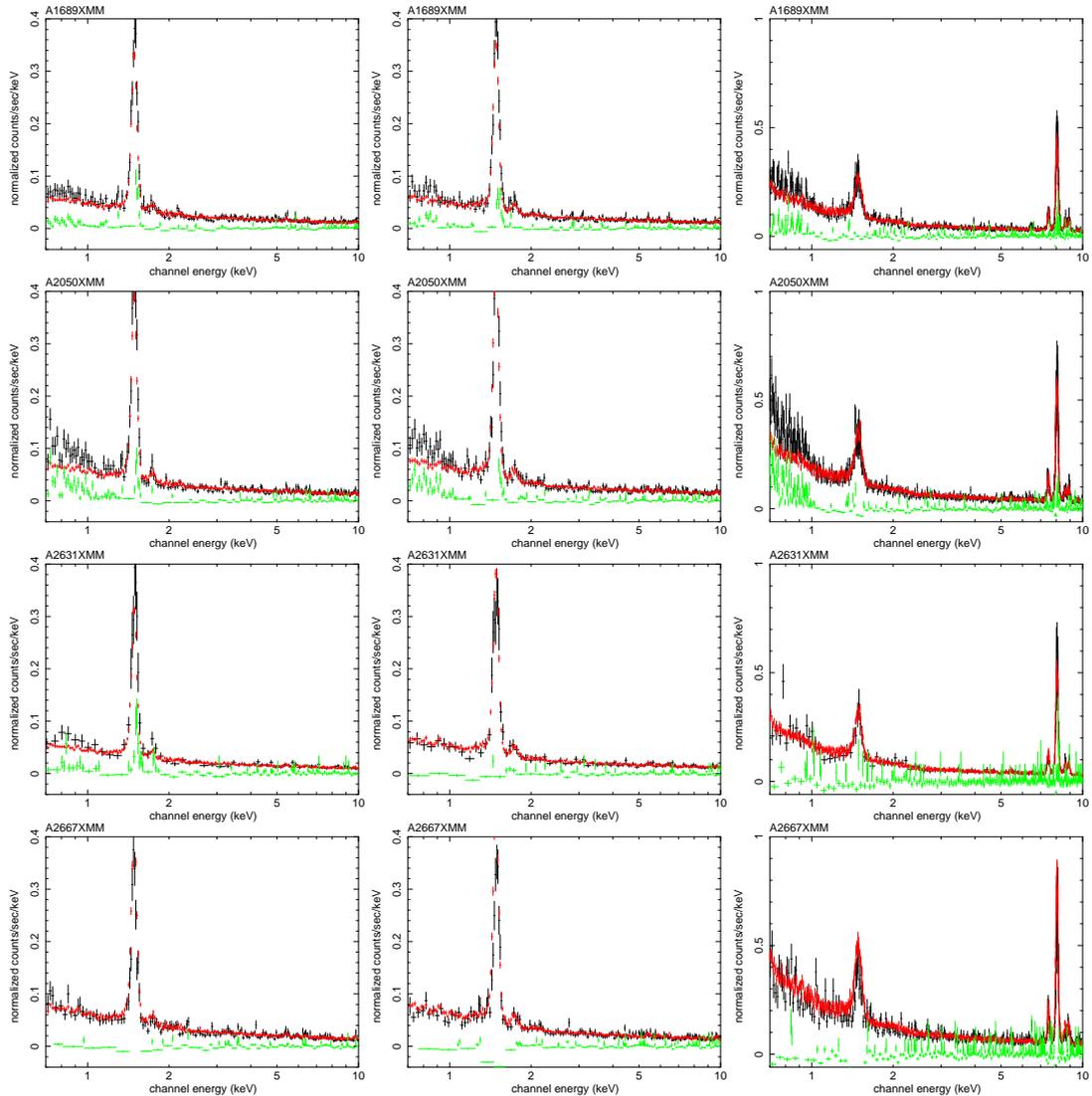

\centering
\includegraphics[width=0.20\textwidth,angle=-90]{Figs/9168fig1.eps}
\includegraphics[width=0.20\textwidth,angle=-90]{Figs/9168fig2.eps}
\includegraphics[width=0.20\textwidth,angle=-90]{Figs/9168fig3.eps}
\includegraphics[width=0.20\textwidth,angle=-90]{Figs/9168fig4.eps}
\includegraphics[width=0.20\textwidth,angle=-90]{Figs/9168fig5.eps}
\includegraphics[width=0.20\textwidth,angle=-90]{Figs/9168fig6.eps}
\includegraphics[width=0.20\textwidth,angle=-90]{Figs/9168fig7.eps}
\includegraphics[width=0.20\textwidth,angle=-90]{Figs/9168fig8.eps}
\includegraphics[width=0.20\textwidth,angle=-90]{Figs/9168fig9.eps}
\includegraphics[width=0.20\textwidth,angle=-90]{Figs/9168fig10.eps}
\includegraphics[width=0.20\textwidth,angle=-90]{Figs/9168fig11.eps}
\includegraphics[width=0.20\textwidth,angle=-90]{Figs/9168fig12.eps}
\caption{Comparison between observation background and Lumb background for MOS1 (left panel), MOS2  (middle panel) and PN (right panel). The black points correspond to the observation data, the red points to Lumb scaled background and the green points represent the residual.}
\label{bkg}
\end{figure*}

The background was normalized using a spectrum obtained in an annulus (between 9-11 arcmin) where the cluster emission is no longer detected. A normalized spectrum was then subtracted, yielding a residual spectrum as shown in Fig.~\ref{bkg}. 

As A496 is a nearby cluster its emissivity extends up to the outskirts of the image and for this reason we could not apply this technique of background normalization. For this cluster, we subtracted the background without normalizing it. If the background had not been well subtracted we would have had an emission excess at the outskirts. Due to this excess, we would have obtained from the fit of the temperature profile higher temperatures toward the outskirts. Even so, our temperature profile is in agreement with the results obtained by \citet{Tamura01}.

\begin{table}[t!]
\centering
\caption{Observational information of the XMM-Newton data. Column~(1): cluster name; Col.~(2): Filter in which cluster was observed; Col.~(3): total exposure time; Col.~(4,5,6): exposure time after flare cleaning in each detector.}
\begin{tabular}{ccccccc}
\hline\hline
 Cluster & Filter & $t_{\rm exp_{\rm tot}}$ (ks)& \multicolumn{3}{c}{$t_{\rm exp}$ (ks)}  \\
         &   & & MOS1 & MOS2 & PN\\
\hline\noalign{\smallskip}
A496  & Thin1  & 30.87 & 14.44 & 14.54 &  8.40 \\  
A1689 & Thin1  & 39.81 & 34.24 & 34.64 & 27.03 \\
A2050 & Thin1  & 28.41 & 24.01 & 24.64 & 19.87 \\
A2631 & Thin1  & 14.12 & 9.32  & 11.35 & 5.00 \\
A2667 & Medium & 31.16 & 22.03 & 22.17 & 13.88 \\
\hline	  
\end{tabular}
\label{obsinfo}
\end{table}

\subsection{Optical: SDSS and CFHT}
In order to investigate the optical component of A1689, A2050 and A2631, we downloaded from the Sloan Digital Sky Survey Data release 5 \citep{DR5} the model magnitudes dereddened for all galaxies inside $r_{500}$ (derived in this work from an X-ray analysis, see Sect. \ref{massdet}). Since extinction is less in the $i^{\prime}$ band we used this band to construct luminosity functions. The galaxy catalog is essentially complete down to 23.5 $i^{\prime}$ magnitude.

A496 was observed with the Canada-France-Hawaii Telescope (CFHT) with the Megacam camera in 2003 (program 03BF12, P.I. V. Cayatte). Megacam covers a field of 1 $\times$ 1 square degrees with a pixel size of 0.187 arcsec. Images were obtained in the $u^{\ast}$, $g^{\prime}$, $r^{\prime}$ and $i^{\prime}$. 
The Terapix data center at IAP (France) reduced the images, correcting bias and flat field and calibrating photometrically. For more details about data reduction see \citet{Boue07}.

\section{Gas density distribution}\label{gasdist}

\subsection{Surface brightness profile}\label{bs}

The X-ray emissivity of the intra-cluster medium depends strongly on the gas density ($\propto n^{2}$) and weakly on its temperature ($\propto T^{1/2}$), which means that the observed projected emissivity can be used to accurately measure the gas density profile. 

In order to compare theory with observations, a description of the gas distribution is needed. Using a parameter-dependent model for the gas density profile, it is possible to reconstruct the 3D X-ray emission of the cluster.

The surface brightness profile has been described for many years by the 
$\beta$-model \citep{CavFusco76}. Satellites with better spatial resolution showed a significant difference between data points and $\beta$-model at low radii for cool-core clusters \citep{JF84,XueWu2000}. An enhancement of the gas density in the central region leads to cooling flows and, as a consequence, the gas density radial profile is cuspy in the center instead of having a flat core. Following this observational evidence, alternative parametrizations were tried to model this emission excess, notably with the use of the double $\beta$-model \citep{Ikebe99,Vik06,Pratt05}. In this model, a second term was introduced to take into account the change of slope in the density profile in the cluster core. This leads to a six parameter function ($S_{cool}$, $r_{cool}$, $\beta_{cool}$, $S_{amb}$, $r_{amb}$, $\beta_{amb}$). 

Another approach towards the description of the surface brightness profile was carried out by \citet{Pislar97}, who used the S\'{e}rsic model to describe the X-ray surface brightness profile. 

In a more recent work, \citet{Demarco03} presented a detailed analysis of 24 galaxy clusters, showing that the S\'{e}rsic model was adequate to describe the X-ray surface brightness profile. \citet{Durret05} used the S\'{e}rsic description for Abell 85 and compared it to the $\beta$-model: with all points included the S\'{e}rsic law provided the best fit. Moreover, without the inner points, the S\'{e}rsic and $\beta$-model become indistinguishable.

The central part of cluster is the region where most energy is radiated and therefore it is responsible for the majority of the X-ray luminosity. Thus, it is important to adequately model the inner parts of a cluster. 

\begin{figure*}[t!]
\centering
\includegraphics[width=0.60\textwidth,angle=90]{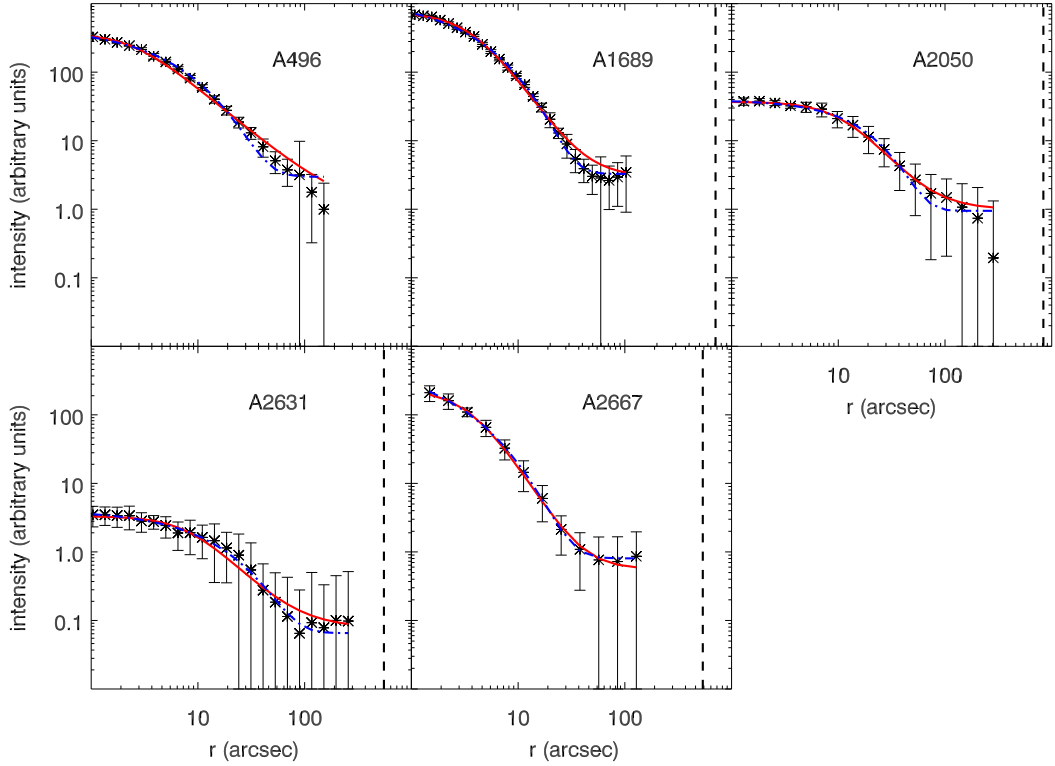}
\caption{Surface brightness profile (black points) of our clusters A496, A2667, A2050, A2631 and A1689. $\beta$-model fits are represented by red continuous lines and the S\'{e}rsic fits are represented by blue dashed lines. The vertical dashed lines represent $r_{500}$. For A496 the $r_{500}$ is beyond the range of the x-axis.}
\label{BSfigs}
\end{figure*}

We used the S\'{e}rsic model and the $\beta$-model to describe the surface brightness profiles of our sample with the fits presented in Fig.\ref{BSfigs}. For $\beta$-model fits the inner points were also included.

For each cluster an image and the corresponding exposure maps were created to correct bad pixels and telescope vignetting. The [0.7-7] keV band is used to derive the surface brightness profile, which was obtained adopting  logarithmic steps between consecutive annuli. Obvious point sources or detector gaps were masked and not taken into account.
We used the STSDAS/IRAF {\it ellipse} task.

\subsection{Gas density profile modeling}
The surface brightness profile observed is the projection of the plasma emissivity and we assume two analytical equations to describe the surface brightness profile: the $\beta$-model and the S\'{e}rsic. The former is defined as follows:

\begin{equation}
\label{BS_beta}
\Sigma(R)= \Sigma_{0}\displaystyle \biggl [1+\biggl(\frac{R}{r_{c}}\biggl)^{2}\displaystyle \biggl]^{-3\beta+0.5},
\end{equation}
where $r_{c}$ is the core radius, $\beta$ is the shape parameter and $\Sigma_{0}$ is the central surface brightness. The latter is defined by:
\begin{equation}
\label{BS_sersic}
\Sigma(R)= \Sigma_{0}\exp{\displaystyle \biggl[-\biggl(\frac{R}{a^{\prime}}\biggl)^{\nu}\displaystyle \biggl]},
\end{equation}
where $a^{\prime}$ is the scale parameter, $\nu$ (usually represented as 1/n) is the shape parameter and $\Sigma_{0}$ is the central surface brightness.

The density profile of the ICM, which emits via thermal bremsstrahlung, is the deprojection of the surface brightness given above and can be given by: 

\begin{equation}
\label{n0_beta}
\rho(r)= \rho_{0}\biggl [1+\biggl(\frac{r}{r_{c}}\biggl)^{2}\biggl]^{\frac{-3\beta}{2}},
\end{equation}
for the $\beta$-model and, for the S\'{e}rsic we have:

\begin{equation}
\label{n0_sersic}
\rho(r)= \rho_{0}\biggl(\frac{r}{a}\biggl)^{-p^{\prime}} \exp\displaystyle \biggl[-\biggl(\frac{r}{a}\biggl)^{\nu}\displaystyle \biggl],
\end{equation}
where $p^{\prime}=p/2$, $p=1-0.6097 \nu +0.05563 \nu^{2}$ and $a=a^{\prime}~2^{1/\nu}$ \citep{Durret05}.

Note that from the mathematical point of view, the S\'{e}rsic has fewer parameters than the double $\beta$-model and, we can adequately compute important quantities such as the total mass with no more than three parameters.

For cool-core clusters, the temperature profile decreases towards the center and, if one assumes that in this inner region the derivative of the temperature as a function of the radius is positive and that the surface brightness profile is described by the $\beta$-model we would obtain a negative mass in the central part. Mathematically, we have:
\begin{equation}\label{mtot}
M(r) \propto - r \left(\frac{\mathrm{d} \; \ln T}{\mathrm{d} \; \ln r} + \frac{\mathrm{d} \;\ln \rho}{\mathrm{d} \; \ln r}\right),
\end{equation}
which gives the dynamical mass of a cluster. Assuming the $\beta$-model, we have: 
\begin{equation}
\lim_{r \to 0} \;\; \frac{\mathrm{d} \;\ln \rho}{\mathrm{d} \; \ln r} = 0.
\end{equation}
For $r \ll r_{c}$, $T \propto r^{\alpha}$ and:
\begin{equation}
\lim_{r \to 0} \;\; \frac{\mathrm{d} \;\ln T}{\mathrm{d} \; \ln r} = \alpha.
\end{equation}
These lead to:
\begin{equation}
M(r) \to -\frac{k \;T \;r \;\alpha}{G \; m_{p} \; \mu} < 0,
\end{equation}
which is not a physical solution.

\begin{table*}[t!]
\centering
\caption{Fits results. Column~(1): cluster name; Col.~(2): slope parameter of the $\beta$-model; Col.~(3,4): scale radius in arcsec and in $h_{\rm 70}^{-1}$ kpc of the $\beta$-model; Col.~(5): electron number density of the central region calculated by means of the $\beta$-model; Col.~(6): slope parameter of the S\'{e}rsic fit; Col.~(7,8) scale radius in arcsec and in $h_{\rm 70}^{-1}$ kpc of the S\'{e}rsic model; Col.~{9} electron number density of the central region calculated by means of the S\'{e}rsic model.}
\begin{tabular}{c| c c c c|c c c c}
\hline\hline
&\multicolumn{4}{c|}{$\beta$-model}& \multicolumn{4}{c}{S\'{e}rsic} \\
\cline{2-9}
Cluster & $\beta$ & $r_{c}$ &  $r_{c}$ & $n_{0}$& $\nu$ & a$^{\prime}$ & a$^{\prime}$& $n_{0}$ \\
         &             & (arcsec)    & ($h_{\rm 70}^{-1}$ kpc)   & $\rm cm^{-3}$   &  & (arcsec)  & ($h_{\rm 70}^{-1}$ kpc) & $\rm cm^{-3}$   \\
\hline
A496  & 0.410  $\pm$ 0.015 & 23.48  $\pm$ 0.28 & 15.48 $\pm$ 0.18 & 0.084& 0.390  $\pm$ 0.046 & 3.6  $\pm$ 0.23   & 2.37   $\pm$ 0.15  &0.036 \\
A1689 & 0.550  $\pm$ 0.025 & 25.72  $\pm$ 0.23 & 79.16 $\pm$ 0.71 & 0.065 & 0.650  $\pm$ 0.030  & 12.78 $\pm$ 0.16 & 39.32 $\pm$ 0.48 & 0.012 \\
A2050 & 0.43   $\pm$ 0.028  & 10.22   $\pm$ 0.73  & 22.0 $\pm$ 1.5 & 0.026 & 1.285 $\pm$ 0.047  & 15.33   $\pm$ 0.32    & 56.4  $\pm$ 1.2 & 0.010\\
A2631 & 0.732  $\pm$ 0.073  & 21.2   $\pm$  2.5$\pm$   & 82.6    $\pm$ 9.6  &  0.051 & 1.001  $\pm$ 0.066  & 14.9  $\pm$ 1.2   & 58.0  $\pm$ 4.7 & 0.013 \\
A2667 & 0.660  $\pm$ 0.017 & 30.8  $\pm$ 1.9 & 113.25  $\pm$ 6.99  & 0.039 & 0.570 $\pm$ 0.034 & 6.3    $\pm$ 1.5   & 23.1   $\pm$  5.4 & 0.022\\
\hline	
\end{tabular}  
\label{fitsinfo}
\end{table*}

\section{Temperature distribution}\label{tempdist}

\subsection{Temperature profile: spectral analysis}\label{tempprof}

We extract cluster spectra in concentric annuli around the cluster center. On one hand, each annuli had to have enough source counts to have good signal-to-noise ratio ($S/N$). On the other hand, the regions must be small enough to determine the temperature profile in small bins. We excluded from our analysis the most luminous point sources which were located inside the FOV of the cameras. Each annulus had at least 1200 net counts (i.e., background subtracted), which corresponds to a $S/N$ of at least 15.

For the spectral fit, we used XSPEC version 11.0.1 \citep{KArnaud_xspec} and modeled the obtained spectra with a Mekal single temperature plasma emission model \citep{KM93,Lied95}. The free parameters are the X-ray temperature ({\it kT}) and the metal abundance (metallicity). Spectral fits were done in the energy interval of [0.7-8.0] keV and, with the hydrogen column density fixed at the galactic value (see Table~\ref{generalinfo}) estimated with the task {\it nH} of FTOOLS \citep[based on][]{DL}

\subsection{Temperature profile modeling and deprojection}

\label{tempmod}
We have chosen an empirical function that can describe the overall temperature profiles of cool-core and non cool-core clusters: a decrease in temperature towards the cluster center, a maximum value and a decrease or a flat profile towards the outskirts. The function is given by Eq.~\ref{eqtemp}: 

\begin{equation}
\label{eqtemp}
T_{2D}(R)=\frac{T_{0} \; \displaystyle \bigg [\alpha \; \sqrt{\frac{R}{r_{t}}}+ \frac{R}{r_{t}}+1\displaystyle \bigg]}{\biggl(\frac{R}{r_{t}}\biggl)^{2}+1},
\end{equation}
where $r_{t}$ is the scale parameter, $T_{0}$ is the central temperature and $\alpha$ is the shape parameter.

The difference between the deprojected temperature profile (3D) and the projected (observed) temperature profile (2D) fitted above should be less than $10\%$, within the observational uncertainties \citep{Mark99,Komatsu01}.
To be more realistic, we deprojected the temperature profile as described in \citet{Limaneto06} to obtain the 3D temperature profile given by Eq.~\ref{eqtempdeproj}:  

\begin{equation}
\label{eqtempdeproj}
T_{3D}(r)=\frac{\int_{r}^{\infty}\displaystyle \bigg(\frac{\partial[\Sigma '(R)T_{2D}(R)]}{\partial R} \frac{dR}{\sqrt{R^{2}-r^{2}}}\displaystyle \bigg) }{\int_{r}^{\infty} \displaystyle \bigg(\frac{\partial \Sigma '(R)}{\partial R} \displaystyle \bigg)\frac{dR}{\sqrt{R^{2}-r^{2}}}},
\end{equation}
which will be used in the determination of the dynamical masses. For non cool-core clusters we simply used the mean temperature to infer the total masses.

To compare the properties of these clusters it is useful to determine a representative temperature as the global mean temperature. The mean temperature was determined extracting cluster spectra inside an external annulus which encompasses the majority of the X-ray data. For each cluster the value of the mean temperature is given in Table~\ref{generalinfo}.

\begin{figure}[t!]
\centering
\includegraphics[width=0.22\textwidth,angle=90]{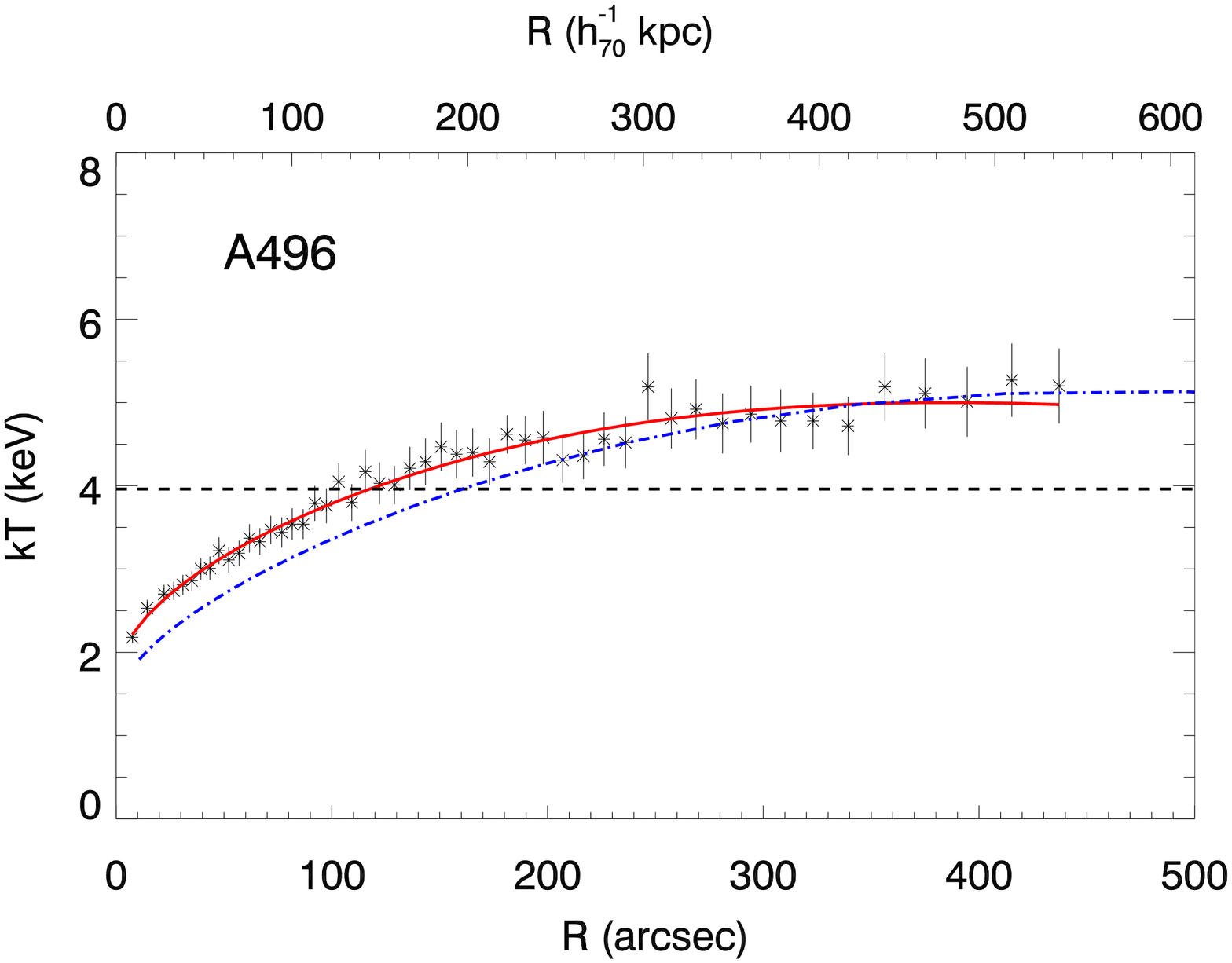}
\includegraphics[width=0.22\textwidth,angle=90]{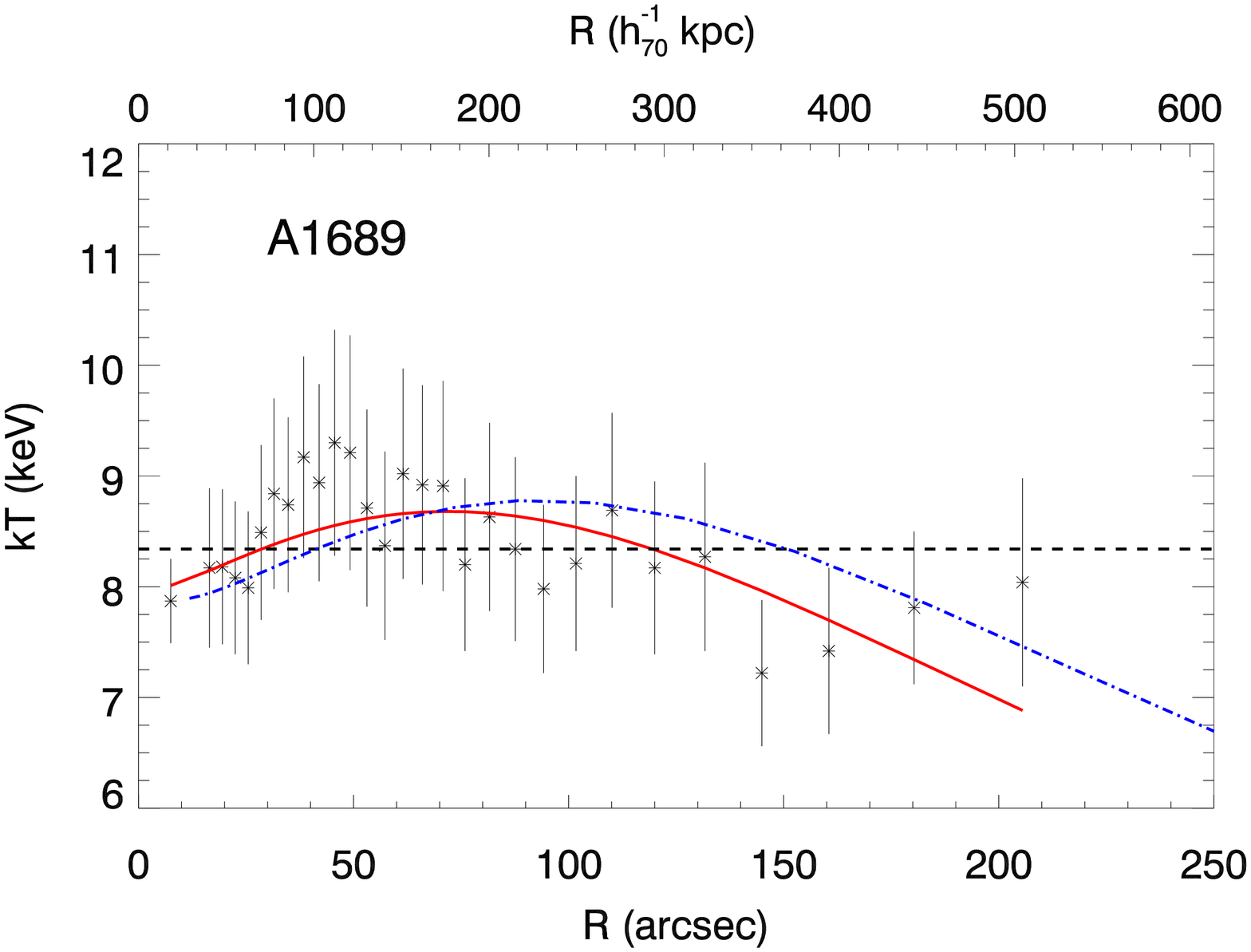}
\includegraphics[width=0.22\textwidth,angle=90]{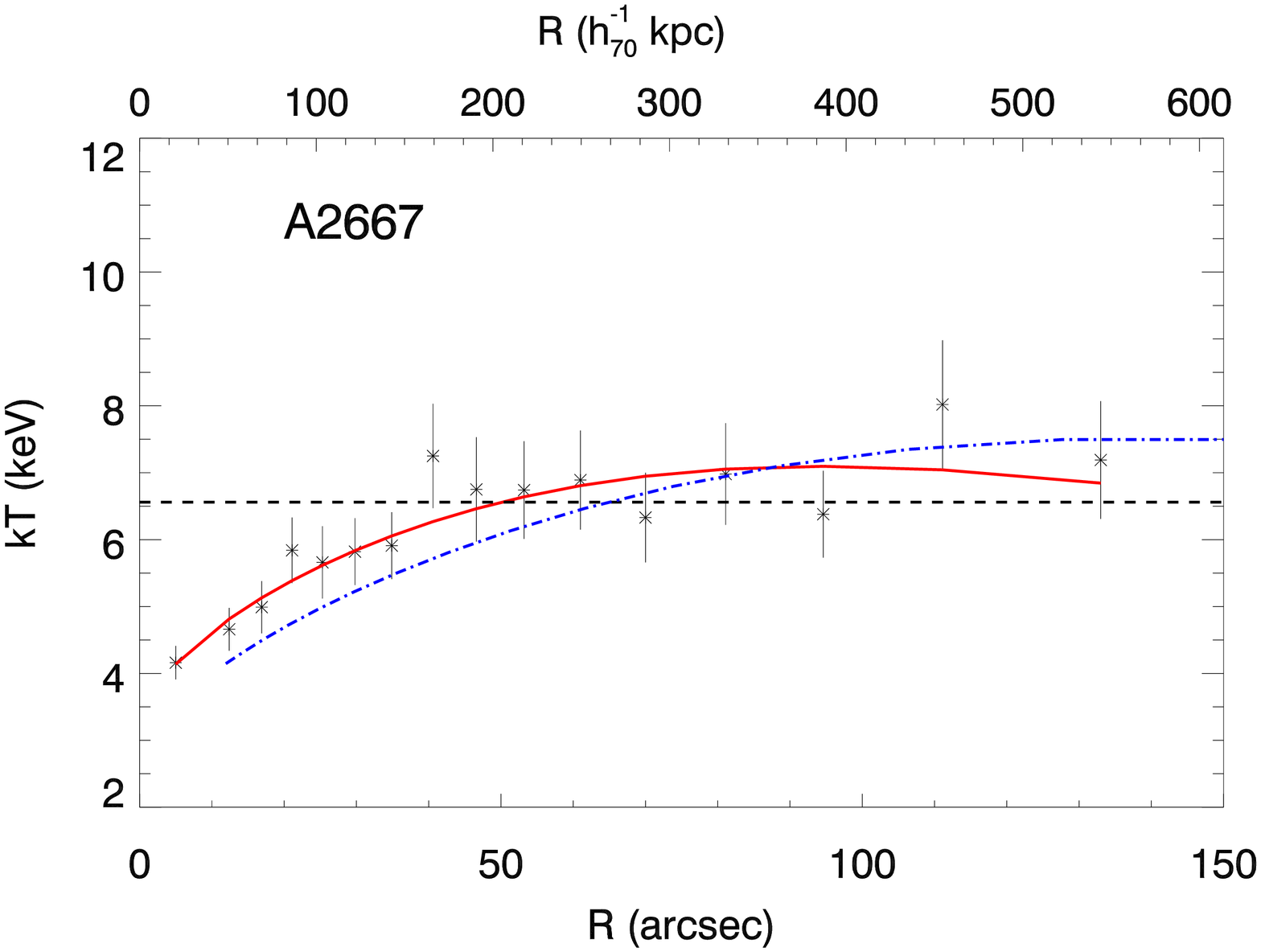}
\includegraphics[width=0.22\textwidth,angle=90]{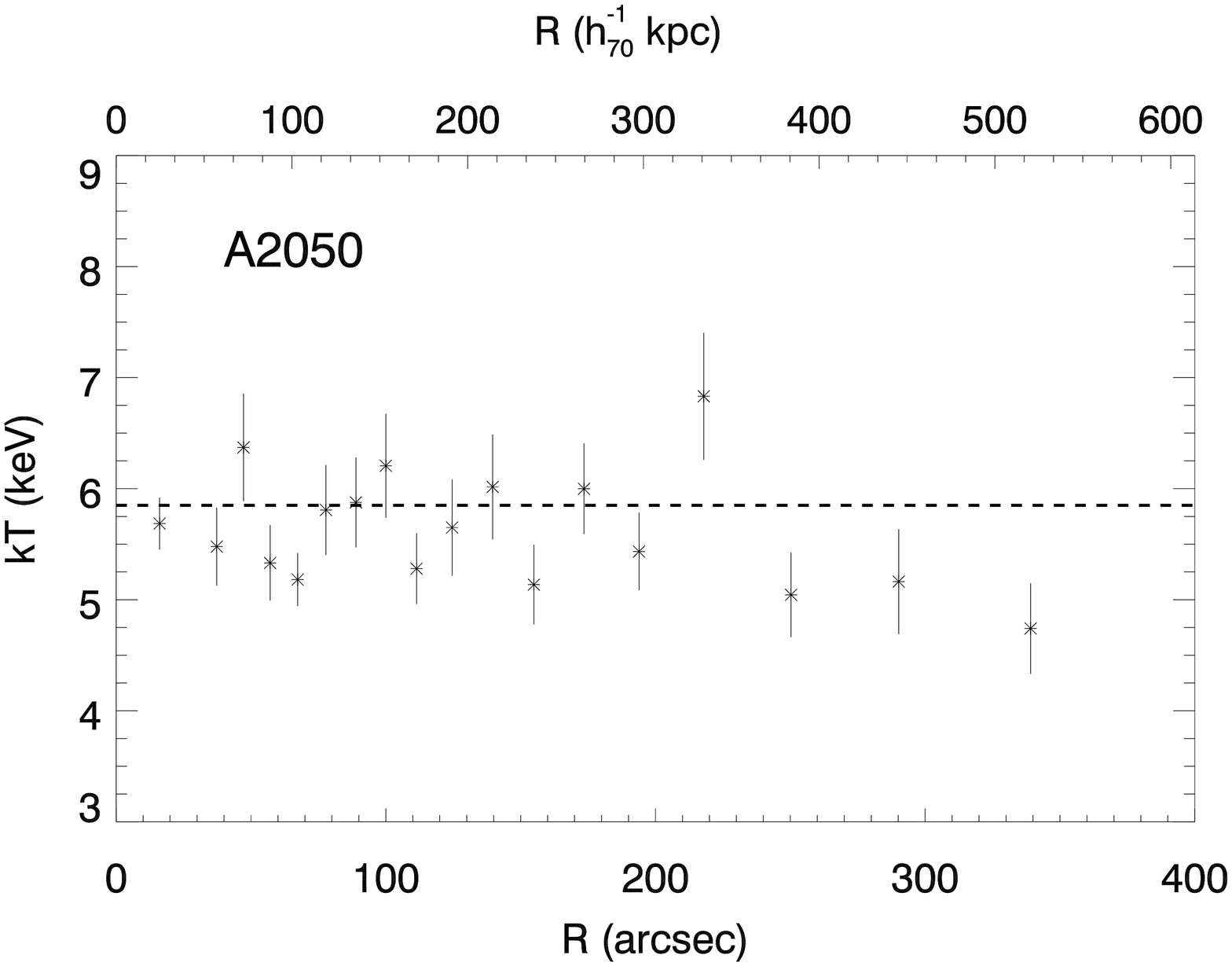}
\includegraphics[width=0.22\textwidth,angle=90]{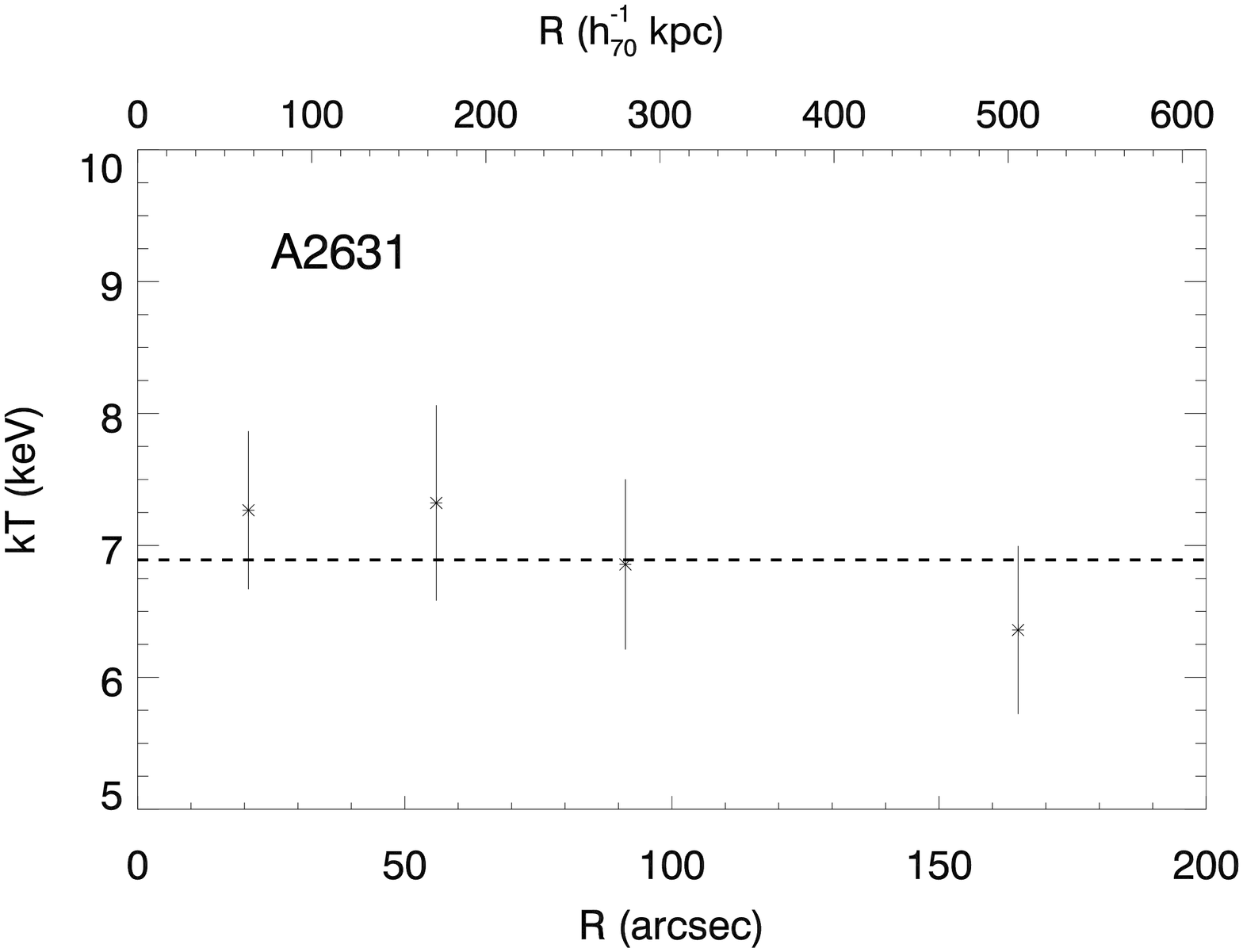}
\caption{Temperature profile for A469, A1689, A2667, A2050 and A2667. For A496, A1689 and A2667 2D analytical fits given by Eq.~\ref{eqtemp} are represented by the red solid lines, the blue dot-dashed lines represent the computed deprojected temperature profile given by Eq.~\ref{eqtempdeproj} and the mean temperatures are represented by the horizontal dashed lines. For A1689, A2050 and A2631 the mean temperatures are represented by the horizontal dashed line.}
\label{AjusteTemp}
\end{figure}

In Fig.~\ref{AjusteTemp} we present temperature profiles with 2D analytical fits of Eq.~\ref{eqtemp}, 3D deprojected temperature profiles given by Eq.~\ref{eqtempdeproj} and mean temperatures for clusters in our sample.

\begin{table}[t!]
\centering
\caption{Best fit results for the parameters of 2D temperature profiles (Eq.~\ref{eqtemp}) for A496, A1689 and A2667.}
\begin{tabular}{cccc}
\hline\hline
 Cluster & $T_{0}$ & $r_{t}$ & $\alpha$  \\
         & (keV)  & (arcsec)& \\
\hline\noalign{\smallskip}
A496  & 1.658 $\pm$ 0.069 & 726 $\pm$ 80 & 3.20 $\pm$ 0.24 \\
A1689 &  8.16 $\pm$ 0.58 &  200 $\pm$ 26 & -0.27 $\pm$ 0.17 \\
A2667 &  3.21 $\pm$ 0.39 &  184 $\pm$ 44 & 1.96 $\pm$ 0.53 \\
\hline	  
\end{tabular}
\label{ParamsTemp}
\end{table}

\section{Dynamical analysis}
\label{dynana}

The X-ray morphology can give interesting qualitative insights into the dynamical status of a given cluster \citep{JF84}. Nevertheless, to quantitatively characterize the dynamical state, density and temperature distributions are required. 

During a merger process, the ICM is compressed and consequently the local density and temperature increase  \citep[see, e.g.,][]{Vikhlinin01}.  
As X-ray substructures depend on surface brightness, temperature maps and X-ray substructures are auxiliary tools in the analysis of the dynamical state of clusters \citep[see, e.g.,][]{Durret05}.

Star formation efficiency in galaxies is related to the dynamical state of
galaxy clusters. We present in Fig.~\ref{mapaTcontSE} temperature maps and
respective X-ray substructures overlaid with the X-ray intensity contours.

\subsection{Temperature maps}\label{tempmaps}

The temperature maps are made in a grid, where in each pixel we have made a spectral fit (assuming bremsstrahlung + line emission) to determine the temperature. We set a minimum count number (1200 net counts) for a spectral fit with a MEKAL plasma model.
If we do not reach this minimum count number in a pixel we gradually enlarge the area up to a 5 $\times$ 5 pixel region (or 128$^{\prime \prime}$ $\times$ 128$^{\prime \prime}$ region). The best fit temperature value is attributed to the central pixel. When there are not enough counts in the region, the pixel is ignored and we proceed to the next one. We compute the effective area files (ARFs) and the response matrices (RMFs) for each region in the grid. This procedure is presented in \citet{Durret05}.

We see from Fig.~\ref{mapaTcontSE} that temperature maps present a level of detail that is lost in the projected temperature profiles. For example, hotter blobs are evident even in cool-core clusters, which can be globally relaxed clusters. Since shock fronts compress and heat the gas during the shocks, temperature substructures can reveal possible histories of cluster formation, as discussed in \citet{Covone06mapa}.

\subsection{X-ray substructures}
\label{substruc}

\begin{figure*}[!htb]
\centering
\includegraphics[width=0.65\textwidth]{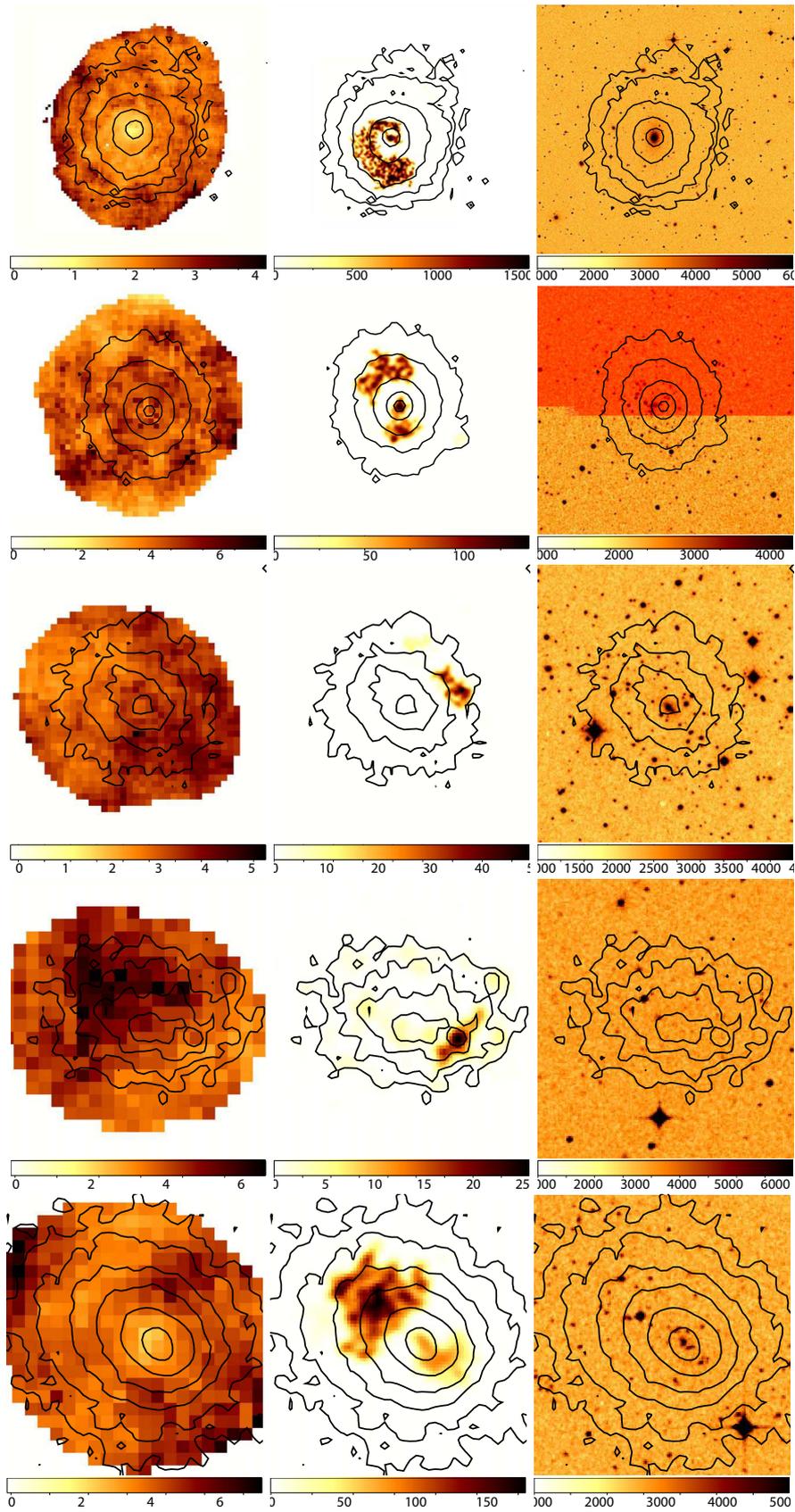}
\caption{Temperature maps, substructure maps and optical image overlaid  
with X-ray contours for A496, A1689, A2050, A2631 and A2667, respectively.}
\label{mapaTcontSE}
\end{figure*}

In Fig.~\ref{mapaTcontSE} we present the images of the X-ray
substructures, which were obtained from the residual images of X-ray $\beta$-model surface
brightness profile fits. The residual image is processed using
percolation and a filter set to be equal to the 3 times the background 
standard deviation. Only pixels with $3 \sigma$ values above the background are
going to be taken into account to determine the positive (emission excess) substructure maps. 

A496 appears to be a very relaxed cluster from its X-ray emission contours. However, its temperature map shows small regions of substructures, with cooler gas seen near the center. This region coincides with the X-ray substructures and also with the cold front found by \citet{Dupke07}. This is evidence of dynamical fossils left after a minor merger process.

Comparing the positive substructure map with the optical image in Fig.~\ref{mapaTcontSE}, we see that for A2050 and A2631 there is a spatial coincidence between the substructures and the bright galaxies, indicating a possible emission peak due to star formation heating the surrounding ICM. However, comparing the positive substructure map with the temperature map, we see that for A496 and A2667 the positive substructures coincide with the low temperature regions. This may be explained by gas infall due to the pressure gradients caused by temperature differences, consequently increasing the density leading to rise in the emissivity. The case of A1689 does not fit in any of the 2 previously discussed cases, and its overdensity may simply be due to the past motion of halos that compressed the ICM, increasing the emissivity.

From substructure maps we see that all clusters but A2667 present evidence of minor mergers. A2667 could have experienced a major merger as proposed by \citet{Covone06mapa}. Even clusters that have a relatively smooth appearance may have undergone several merger events \citep{DLN07}, as revealed by X-ray temperature and substructure maps. The deviation from the equilibrium hypothesis is not significant for the total mass estimates \citep[see, e.g.,][]{Rasia06}.

We see that comparing substructure and temperature maps with optical images gives us hints about the effects of galaxies on the ICM and the effects of ICM motion on temperature. In order to investigate any correlation between star formation efficiency and cluster dynamical state we would need a larger sample, given the intrinsic scatter expected for such quantities.

\section{Stellar baryon content}
\label{stelcont}

We can derive the stellar baryon budget by integrating the luminosity function and using an appropriate stellar mass-to-light ratio. 

The galaxy luminosity function (GLF) is one of the basic statistical properties of galaxies in clusters. The GLF gives the number of galaxies per bin of magnitude and in most cases, the bright galaxies can be fit by a Schechter function \citep{Schechter}. In the case of deep observations there appear to be two components in the GLF: the bright and faint galaxies. This faint end can be seen for magnitudes greater than $M_{i^{\prime}}=18$ and for these cases it is impossible to well describe the GLF with a single Schechter function \citep{Biviano05,Durret02}. 

We have performed analytical fits to the GLF for four (A496, A1689, A2050 and
A2631) of our five clusters in order to estimate the total luminosity and then
the stellar baryon content. For A2667 we used $L_{K}=(1.2 \pm 0.1) \times
10^{12} h_{\rm 70}^{-2} L_{\odot}$ given by \citet{Covone06LK}, based on ISAAC
data \citep{res_isaac}. As this luminosity was computed within
$R=110h_{70}^{-1}\,$kpc in the K band, a filter transformation to the Sloan
$i^{\prime}$ band was done (as described in Sec.~\ref{a2667lk}) and the total
luminosity function was extrapolated to $r_{500}$ as described in
Sect.~\ref{a2667lk}.

\begin{figure}[t!]
\centering
\includegraphics[width=0.32\textwidth,angle=90]{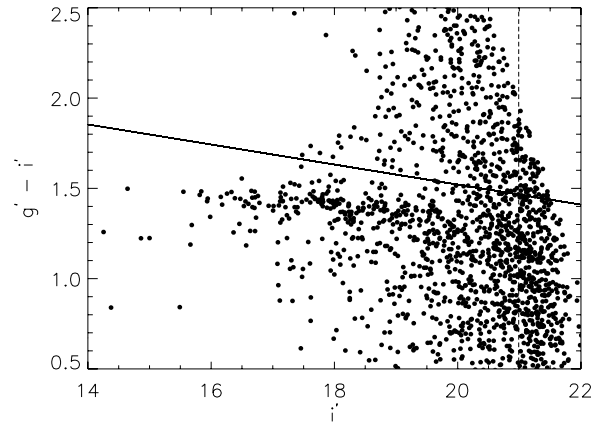}
\caption{Color-magnitude diagram of galaxies within $r_{500}$ in the field of view of A2050. The near-horizontal solid line indicates the separation between cluster and background galaxies. The equation of the solid line is $g^{\prime}-i^{\prime}=2.37-0.055i^{\prime}$. The vertical dashed line indicates the $i^{\prime}$ magnitude limit.}
\label{rs}
\end{figure}

We calculated the galaxy luminosity function of each cluster by means of statistical subtraction. In order to reduce the errors, instead of using all the galaxies we used colors to exclude obvious background objects. In Fig.~\ref{rs} we show a $(g^{\prime}-i^{\prime}) \times i^{\prime}$ color magnitude diagram of the cluster A2050 showing all galaxies within $r_{500}$. The prominent near-horizontal ``sequence'' is composed predominantly of elliptical and lenticular cluster members (cluster red sequence, CRS). For each clusters, we linearly fit the CRS ($(g^{\prime}-i^{\prime}) =A + B \times i^{\prime}$) and we discarded all galaxies redder than $(g^{\prime}-i^{\prime}) = A + 0.3~\rm mag$, represented by the solid line in Fig.~\ref{rs}, since those are likely background objects.

The limit in magnitude for each cluster was essentially the magnitude where the counts turn over. All galaxies that satisfied these two criteria were considered to belong to the cluster.

\subsection{Measure of background counts}
\begin{figure}[t!]
\centering
\includegraphics[width=0.32\textwidth,angle=90]{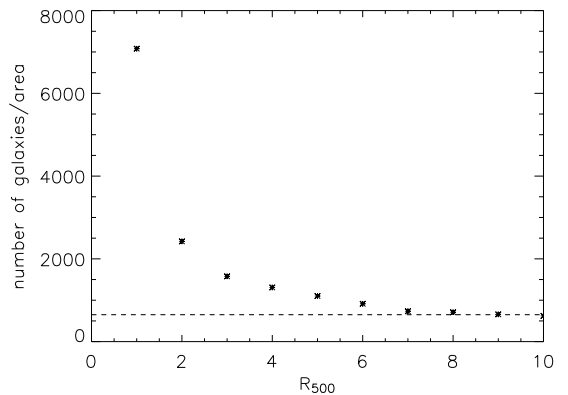}
\caption{Number of galaxies per area of each annulus as a function of $r_{500}$ for A1689. The background can be determined beyond $R= 7 r_{500}$.}
\label{counts_area}
\end{figure}

In order to accurately compute the cluster luminosity function we needed to statistically subtract the background from galaxy counts in the cluster direction.
Field counts should be computed in a region far enough from the cluster for it not to be contaminated by cluster galaxies. 
In Fig.~\ref{counts_area} we show how the number of galaxies per annulus area decreases with radii. From $R = 7~r_{500}$ we see that galaxy counts becomes flat, meaning that the cluster contribution is no longer significant and that these counts are only due to the background. For this purpose, we selected an external annulus whose internal and external radii are $8~r_{500}$ and $9~r_{500}$, where we measure the field counts. 

For the number count of galaxies, we have:
\begin{equation}
N_{\rm cl}(m)=N_{\rm cl+bkg}(m)-\gamma N_{\rm bkg}(m),
\end{equation}
where $N_{\rm cl}(m)$ is the number of cluster galaxies at a certain magnitude {\it m}, $N_{\rm cl+bkg}(m)$ is the total number of galaxies in the line of sight at a certain magnitude, $N_{\rm bkg}(m)$ is the number of background galaxies at the same magnitude and $\gamma$ is the ratio between cluster and control field areas.
After subtracting the background, cluster counts per bin of magnitude in the $i ^{'}$ band were determined.

\subsection{The luminosity functions}

Fitting cluster counts per bin of magnitude with a Schechter function described by:

\begin{equation}
\label{SingleSchech}
\phi(m) dm=\phi^{*} \; 10^{0.4(m^{*}-m)(\alpha+1)} \; \exp{[-10^{0.4(m^{*}-m)}]}
\end{equation}
we obtained the values listed in Tab.~\ref{fitFdL}, where $m^{*}$ is the characteristic magnitude and $\alpha$ is the slope of the LF at faint magnitudes.

Schechter function fits were performed using an IDL code based on the {\it curvefit} function.

\begin{figure}[t!]
\centering
\includegraphics[width=0.25\textwidth,angle=90]{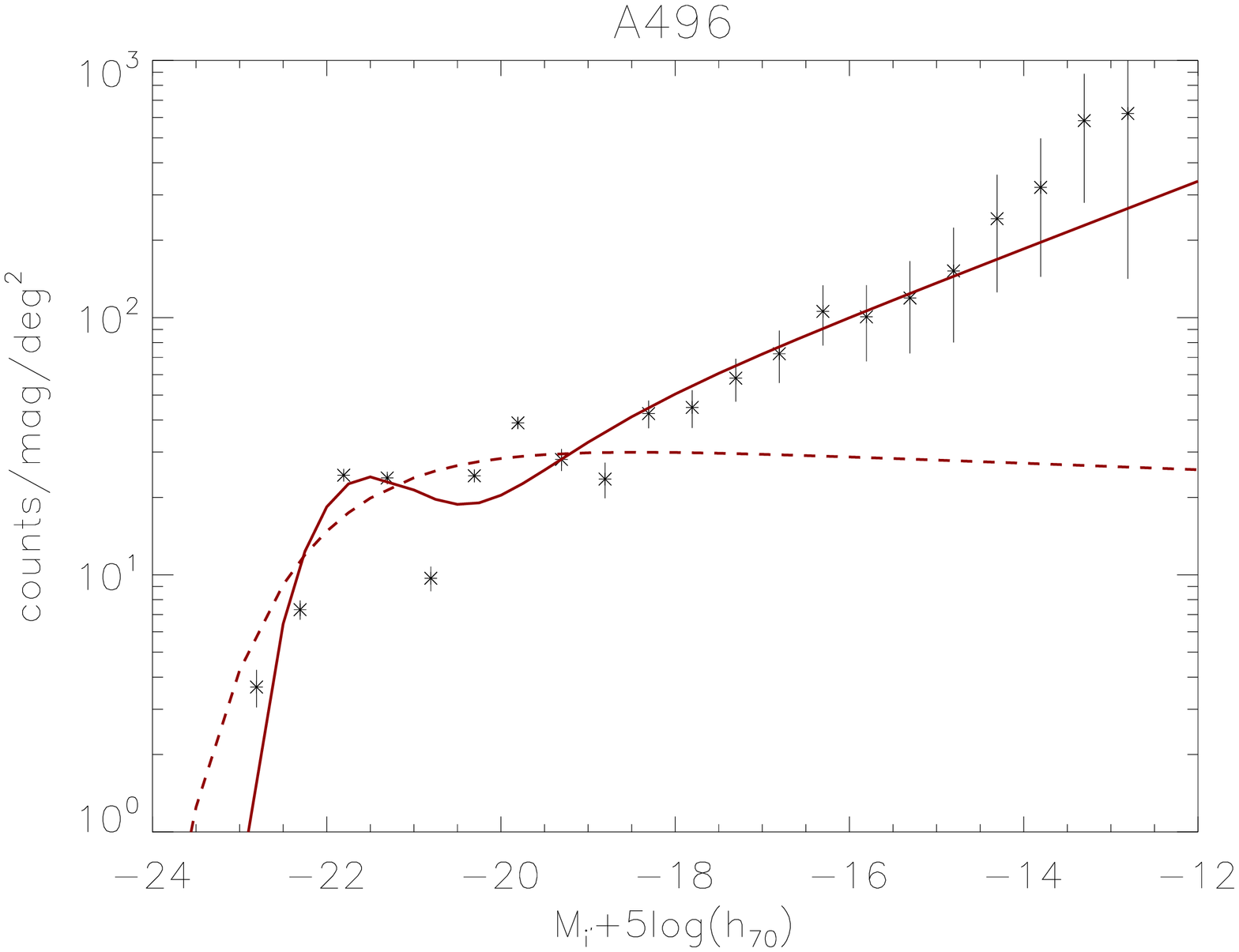}
\includegraphics[width=0.25\textwidth,angle=90]{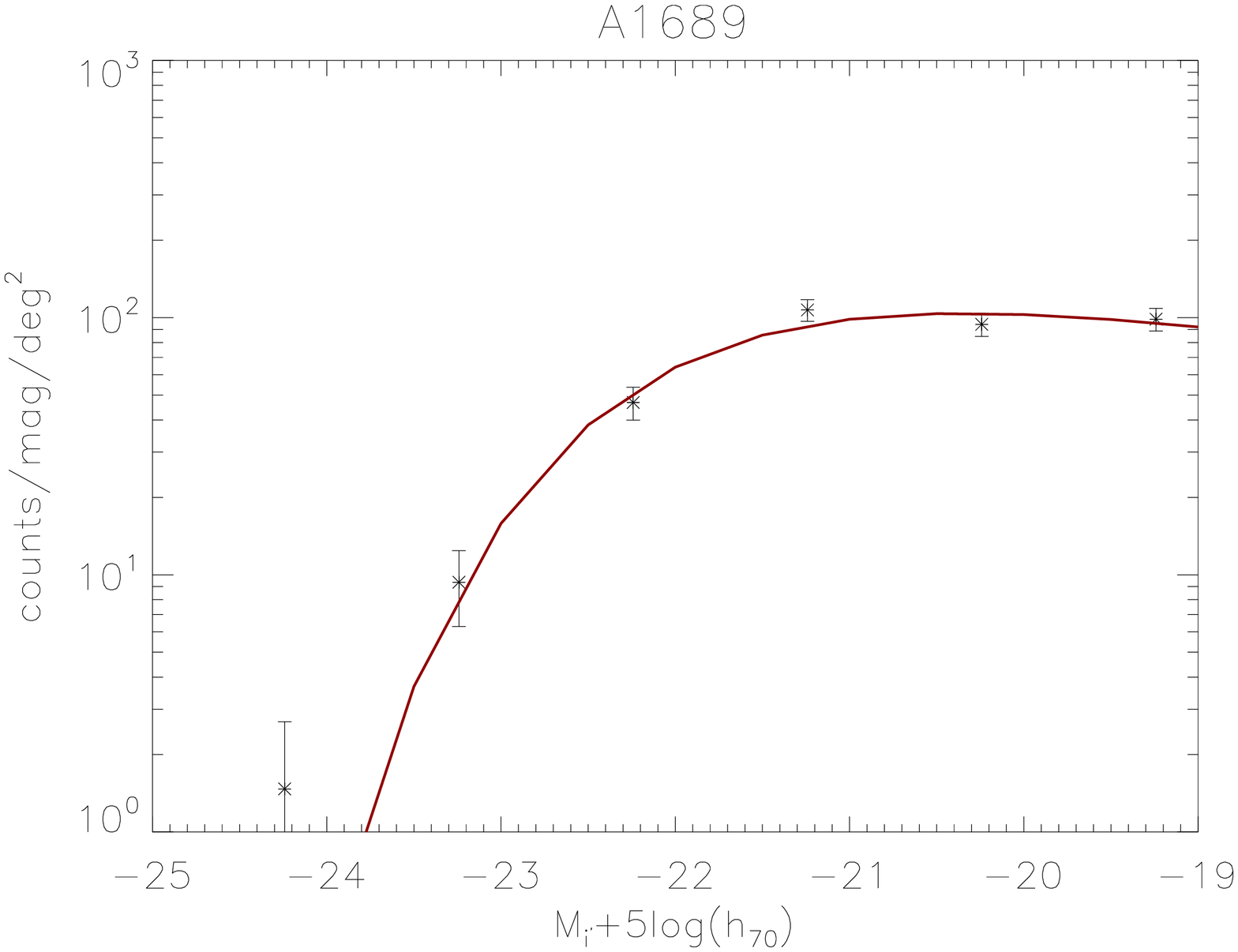}
\includegraphics[width=0.25\textwidth,angle=90]{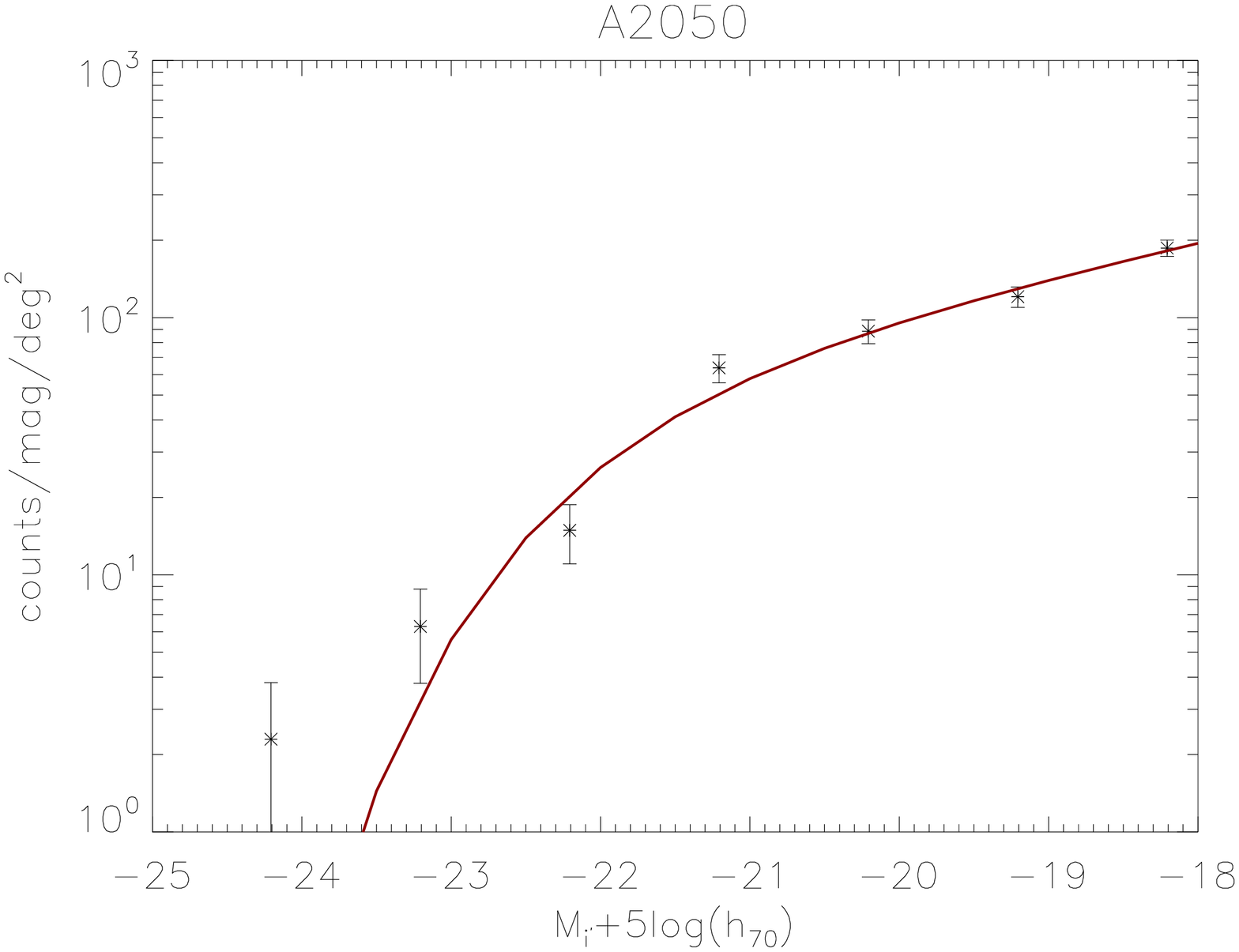}
\includegraphics[width=0.25\textwidth,angle=90]{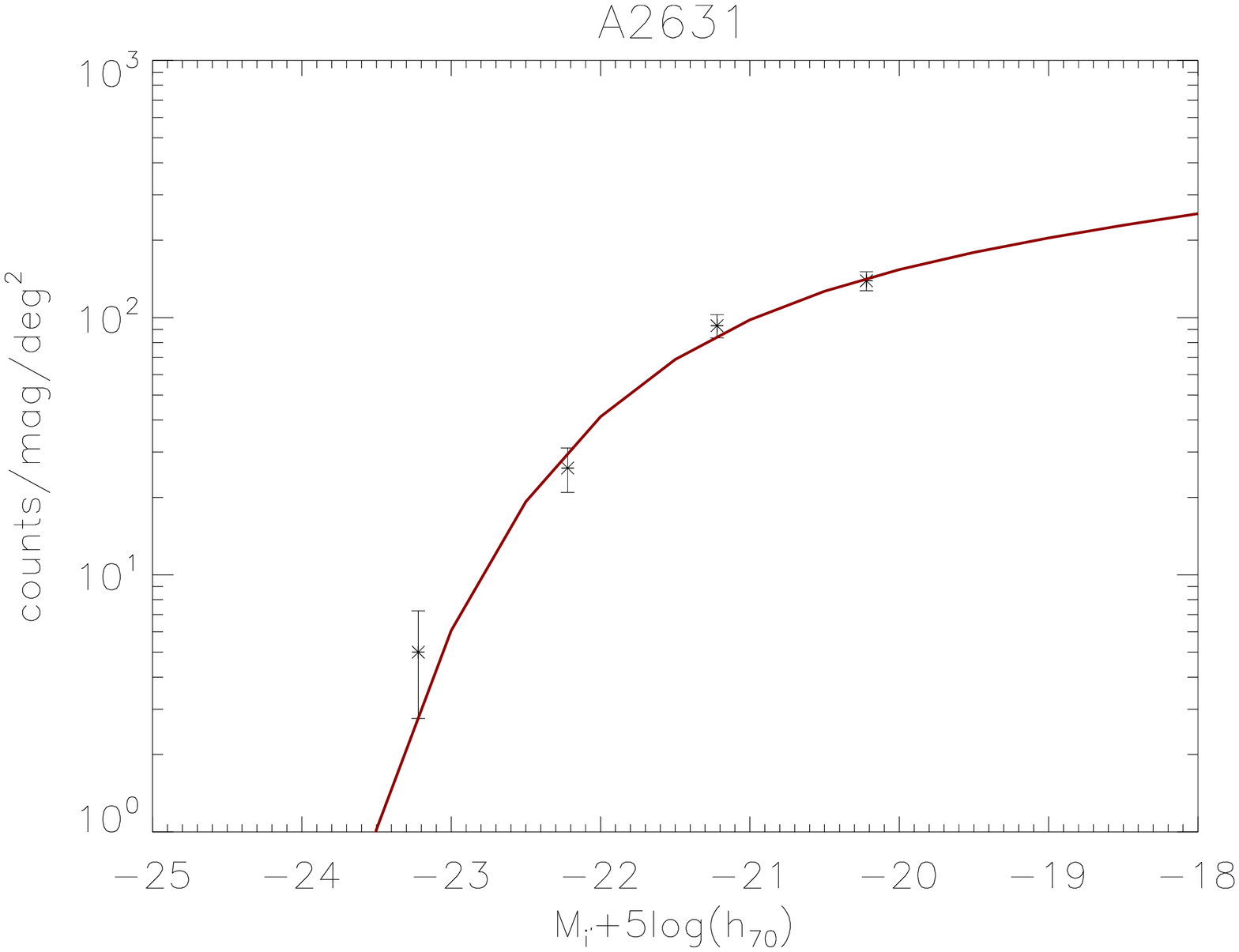}
\caption{Luminosity function for clusters for which we have optical data: A496, A1689, A2050 and A2631, respectively. For A496 the double Schechter fit (given by Eq.~\ref{DoubleSchech}) is represented by the solid line and the single Schechter fit (given by Eq.~\ref{SingleSchech}) is represented by the dotted line. For all other clusters we present only the single Schechter fit.}
\label{LFfit}
\end{figure}

\begin{table*}[t!]
\caption{Luminosity function fits results. For A496 we present the results for single and double Schechter fits.}
\centering
\begin{tabular}{c|ccc}
\hline\hline
& \multicolumn{3}{c}{Schechter} \\
\cline{2-4} 
Cluster  & $\alpha$ & $M_{i}^{\ast}$ & $\alpha_{2}$   \\
\hline
A496  &  -0.968 $\pm$ 0.038 & -22.185 $\pm$ 0.099 & -  \\
     &   1.580 $\pm$ 0.055   & -20.551   $\pm$   0.035 & -3.908 $\pm$ 0.076 \\
\hline
A1689 &   -0.76 $\pm$  0.10 &  -21.93 $\pm$  0.19 & -   \\
\hline
A2050 &    -1.328  $\pm$  0.072 & -22.22 $\pm$  0.31& -  \\
\hline
A2631 &     -1.19  $\pm$ 0.23 & -21.86 $\pm$  0.38 & - \\
\hline
\end{tabular}
\label{fitFdL}
\end{table*}

To obtain $m^{*}$ in absolute magnitude we transform the apparent magnitude to absolute magnitude according to:

\begin{equation}
M^{*}=m^{*} - 25 - 5 \log(d_{\rm L}/ \rm 1 Mpc) + \epsilon(z) - K(z),
\end{equation}
where $d_{L}$ is the luminosity distance, $\epsilon(z)$ is the correction for evolution and K(z) is the K-correction. Assuming that the early-type galaxies are the main population for our galaxy clusters at the cluster redshift we used K-correction values for early-types supplied by \citet{Poggianti97}.

For all clusters except A496 we could comfortably fit the luminosity functions using a single Schechter function. For A496 however, we had to add another power-law component:

\begin{eqnarray}\label{DoubleSchech}
\lefteqn{ \phi(m) dm = \phi^{*}  10^{0.4(m^{*}-m)(\alpha+1)} \; \exp{[10^{0.4(m^{*}-m)}]} \times {} }
\nonumber \\
& & {} \ \ \ \ \ \ \ \ \ \ \bigg[1 + 10^{0.4({m}^{*}-m)(\alpha_{2}+1)}\bigg] {},
\end{eqnarray}
where $\alpha$ and $\alpha_{2}$ are the shape parameters of the bright and the faint part of the luminosity function, respectively.

We did this because the deeper CFHT data for this cluster allowed us to sample the faint-end of the luminosity function ($M^{\star} > -18$) which would not be properly fit otherwise.

\subsection{Total luminosity estimation for A2667}
\label{a2667lk}

In order to calculate the Sloan $i^{'}$ band total luminosity we used a
filter transformation from K to $i^{'}$ band. Assuming that the majority of
galaxies are ellipticals we convolved the spectrum of a $z = 0$ early-type galaxy taken from
\citet{col80} shifted to $z=0.23$, with both the K and $i^{\prime}$ filters,
obtaining:

\begin{equation}\label{iK}
({{i}^{'}-K})_{\rm AB} = 1.489.
\end{equation}
Using Eq.~\ref{iK}, the $L_{K}$ given by \citet{Covone06LK} and 
the solar magnitude in the appropriate bands 
\citep{Fukugita95,Allen_AQ} we have $L_{i}=0.98 
\times 10^{12} h_{\rm 70}^{-2} L_{\odot}$ within $110~h_{\rm 70}^{-1}$ kpc.

The total luminosity within $r_{500}$ (see Table~\ref{StarMass}) can be obtained integrating:
\begin{equation}\label{Lum_extrapol}
L=\int_{0}^{r_{500}} j~4 \pi r^{2}~dr, 
\end{equation}
where {\it j} is the luminosity density of galaxy distribution. Assuming a King profile \citep{King} given by:
\begin{equation}
j=j_{0} \bigg[1+{\bigg(\frac{r}{r_{\rm c}}\bigg)}^{2}\bigg]^{-\frac{3}{2}},
\end{equation}
where $r_{\rm c}$ is the core radius, for which we assumed $r_{\rm c}= 170~h ^{-1}$ \citep{G95} and $j_{0}$ is the normalization. Using the ratio between the total luminosities, $L~(R=r_{500})/L~(R=110~h_{\rm 70}^{-1})$, $j_{0}$ vanishes, and $L~(R=r_{500})$ is computed as a function of $L~(R=110~h_{\rm 70}^{-1})$.

\section{Mass determination}
\label{massdet}

To estimate the total baryon fraction from the stellar components and the ICM, we must choose a fixed radius within which the baryonic mass can be computed. We choose $r_{500}$, as it is the largest radius for which the current X-ray data require no model extrapolation \citep{Vik06} and is about the virial radius \citep{LaceyCole93}. 

Baryonic and total masses are determined within this radius, which can be derived according to \citet{LN03}, assuming the isothermal case:
\begin{equation}
r_{\Delta}=r_{c} \bigg[ \frac{4.5 \times 10^{8}~\beta ~\langle kT
\rangle}{\Delta ~h_{70}^{2}~f^{2}(z; \Omega_{m}, \Omega_{\Lambda})~\mu~r_{c}^{2}} \bigg]^{1/2},
\end{equation}
where $\Delta$ is the factor by which the mass density exceeds the critical
density (in this work, $\Delta=500$), $\beta$ and $r_{c}$ are the slope 
and the characteristic radius (given in kpc) given by $\beta$-model fit, 
$\langle kT \rangle$ is the mean temperature given in keV and $f(z; \Omega_{m}, \Omega_{\Lambda})=\Omega_{\Lambda}+\Omega_{m}~(1+z)^3-(\Omega_{m}+\Omega_{\Lambda}-1)~(1+z)^{2}$.

\subsection{Gas and total mass determination from XMM-Newton data}

Combining the gas density and the deprojected temperature profiles, we can derive the total gravitating mass under the assumption of hydrostatic equilibrium and spherical symmetry as shown in Eq.~\ref{mtot}. 

The gas mass is given by:

\begin {equation}
M_{gas}( < r)=\int_{0}^{r_{500}} \rho_{g}~4 \pi{r^{\prime}}^{2}~\mathrm{d} r^{\prime}
\end{equation}
where $\rho_g$ is the gas density given by Eq.~\ref{n0_beta} and Eq.~\ref{n0_sersic}. To estimate the central density ($\rho_{0}= n_{0} m_{p}$) we can use the normalization of the XSPEC model ({\it K} parameter):

\begin{equation}
n_{0}^{2}=\frac{4 \pi D_{A}^{2}(1+z)^2 K 10^{14}}{EI} ~\rm cm^{-6},
\end{equation}
where $D_{A}$ is the angular distance and {\it EI} is the emission integral, which is model dependent. 

The gas and dynamical mass determinations with both models (S\'{e}rsic and $\beta$ models) are presented in Table~\ref{TabMassDet} with the best values in bold.

For clusters that have a $\nu$ value of the S\'{e}rsic fits to the surface brightness profile equal to or greater than 1, the total mass profiles increase rapidly with radius while the gas mass profiles do not have the same behavior. For this reason, the mass determination for A2050 and A2631 is not reliable with the S\'{e}rsic model and so we do not present them in Table~\ref{TabMassDet}.

In order to verify our results, especially for clusters that had the surface brightness profile described by the S\'{e}rsic (A496, A1689 and A2667), we compared our estimates of the gas fraction ($f_{\rm gas}$) presented in Table~\ref{TabMassDet} with literature values. Using a double $\beta$-model to describe the surface brightness profile, \citet{Zhang07} obtained $f_{\rm gas}$ equals $0.102 \pm 0.06$ and $0.128 \pm 0.073$ for A1689 and A2667, respectively. \citet{LMS} obtained $f_{\rm gas}=0.1347 \pm 0.32$ for A496 also with the double $\beta$-model. These results show that despite the difference in methods, the estimates of the gas fraction are in perfect agreement within $r_{500}$.

\subsection{Stellar mass determination from optical data}

To compute the stellar budget of each cluster we used:
\begin{equation}
M_{\star} =  L_{\rm tot} \;\bigg < \bigg ({\frac{M_{\star}}{L_{\star}}\bigg)}_{\rm blue}\times f_{\rm sp} + (1-f_{\rm sp}) \times{\bigg(\frac{M_{\star}}{L_{\star}}\bigg)}_{\rm red} \bigg>,
\end{equation}
where $({\frac{M_{\star}}{L_{\star}})}_{\rm blue}$ is the stellar mass-to-light ratio for late-type galaxies, ${(\frac{M_{\star}}{L_{\star}})}_{\rm red}$ is the stellar mass-to-light ratio for  early-type galaxies and $f_{\rm sp}$ is the late-type fraction.

It has been shown that colors are good estimators of morphological type \citep{Kauff03}. Following \citet{Abilio06}, we adopted $(u-i)=2.68$ to distinguish between early- and late-type galaxies, where the early-type galaxies are the redder ones.

From the luminosity function we determined the total luminosity according to
\begin{equation}
 j_{\rm obs}  = \frac{L_{\rm tot}}{V_{cl}} = \int_{L_{\rm lim}}^{\infty} L \Phi(L)  \;\mathrm{d}L,  
\label{eq_dens_lum}
\end{equation}
where $V_{cl}=V(r_{500})$ is the cluster volume, $j_{\rm obs}$ is the luminosity density, which is the total luminosity inside $V_{cl}$ and $L_{\rm lim}$ is the luminosity corresponding to the $M_{i^{\prime}}=-16$ magnitude. This leads to
\begin{equation}
L_{tot}=V_{cl} L^{\ast}\Gamma[(\alpha+2),(\frac{L_{lim}}{L^{\ast}})],
\end{equation}
where $\Gamma[a,x]$ is the incomplete gamma function.

In order to account for the contribution of fainter cluster members, we extrapolate the GLF by integrating Eq.~\ref{eq_dens_lum} with $L_{lim}=0$, which is reduced to
\begin{equation}
\int_{0}^{\infty} L \Phi(L)  \;\mathrm{d}L = L^{\ast} \Gamma(\alpha + 2).
\end{equation}
Using this extrapolation, the total luminosity values would be on average 5$\%$ greater. Since this extrapolation does not greatly change our prior results, we do not use it.

The late-type and early-type stellar mass-to-light ratios were estimated from \citet{Kauff03} and converted to the $i^{\prime}$-band:

\begin{equation}
(M/L_{i})_{\star}=0.74~ M_{\odot}/L_{\odot},
\end{equation}
for late-types and  
\begin{equation}
(M/L_{i})_{\star}=1.70 ~ M_{\odot}/L_{\odot},
\end{equation}
for early-types.

For A2667 and A496 we do not have enough color information to derive the late-type fraction. In these cases we assumed that $80\%$ are early-types and $20\%$ are late-types to derive a stellar mass-to-light ratio, as these values are the average fraction found for the other three clusters.

Fitting just a Schechter expression for the A496 luminosity function gives a stellar mass of $1.56 \times 10 ^{12} M_{\odot}$ instead of $3.51 \times 10 ^{12} M_{\odot}$ obtained from the fit of Eq.~\ref{DoubleSchech}. For this cluster the faint-end contributes  33$\%$ of the stellar mass. This gives us a clue about the error we are introducing in not considering the faint-end for the other clusters. Our stellar baryon content is underestimated and these faint galaxies can contribute differently from cluster to cluster.

\subsection{Intracluster light contribution}

Many authors have detected a diffuse stellar component \citep{Zibetti05,Covone06mapa,Krick07,Gonzales00,Gonzales07} that 
has possibly been stripped out from galaxies that eventually merged to form the brightest cluster galaxy (BCG) \citep{Murante07} or from the tidal stripping of cluster members \citep{Cypriano06}. The general consensus is that every cluster has an intracluster light (ICL) component \citep{Gonzales07}.

The ICL represents the minor component of the baryonic budget, and its contribution to the total luminosity has been estimated to account for $(6-22)\%$ \citep{Krick07}, $(10-15)\%$ \citep{Rudick06} and $10\%$ \citep{Zibetti07}. 
Another approach was used by \citet{Gonzales07} who considered in their analysis the contribution of the giant BCG and the ICL as a single entity (ICL+BCG) that contributes on average 40$\%$ and 30$\%$ of the total stellar light, within $r_{500}$ and $r_{200}$, respectively.

\citet{Covone06mapa} studied the diffuse light of A2667, but because of the much smaller field-of-view of the WFPC2 data they measured only the cD contribution to the total light.

In order to consider the ICL contribution to the baryon budget we assumed, for the five clusters in our sample, a contribution of $\sim 10\%$ to the total luminosity and we used the early-type galaxy mass-to-light ratio to obtain the ICL mass contribution.

\begin{table*}[t!]
\centering
\caption{Gas and total mass determination from X-ray data. Column~(1): cluster name; Col.~(2,3): gas and total mass calculated by means of the $\beta$-model; Col.~(4,5): gas and total mass calculated by means of the S\'{e}rsic model; Col.~(6): gas fraction ($M_{\rm{gas}}/M{\rm{tot}}$) calculated by means of the best model. Values in bold represent the best model for the cluster. }
\begin{tabular}{c| c c | c c| c}
\hline\hline
&\multicolumn{2}{c|}{$\beta$-model}&\multicolumn{2}{c|}{S\'{e}rsic} \\
\cline{2-5}
 Cluster&  $M_{gas}$ ($10^{14} M_{\odot}$) & $M_{tot}$ ($10^{14} M_{\odot}$) &  $M_{gas}$($10^{14} M_{\odot}$) & $M_{tot}$($10^{14} M_{\odot}$) & $f_{\rm gas}$ \\
        &  ($< r_{500}$) &  ($< r_{500}$) &  ($< r_{500}$) &  ( $ < r_{500}$) &  ( $ < r_{500}$)  \\
\hline
A496  & 1.074 $\pm$ 0.040 & 1.997 $\pm$ 0.012 & {\bf 0.330 $\pm$ 0.040} & {\bf 3.270 $\pm$ 0.19} & 0.10 $\pm$ 0.014\\
A1689 & 3.79 $\pm$ 0.17   &  10.87 $\pm$ 0.20 & {\bf 1.672 $\pm$ 0.080} & {\bf 11.14 $\pm$ 0.46} & 0.150 $\pm$ 0.010\\
A2050 & {\bf 0.568 $\pm$ 0.079} & {\bf 4.974 $\pm$ 0.094} & --  & -- & 0.114 $\pm$ 0.016\\
A2631 & {\bf 0.84 $\pm$ 0.16} & {\bf 11.78 $\pm$ 0.58} &  --  & --  & 0.072 $\pm$ 0.014\\
A2667 & 1.789 $\pm$ 0.047     & 7.06 $\pm$ 0.14 & {\bf 1.132 $\pm$ 0.066} & {\bf 11.63 $\pm$ 0.24} & 0.0973 $\pm$ 0.0060\\
\hline	
\end{tabular} 
\label{TabMassDet}
\end{table*}

\begin{table*}[t!]
\centering
\caption{Results derived from optical data. Column~(1): cluster name; Col.~(2): total luminosity derived from stars within galaxies; Col.~(3): fraction of star forming galaxies; Col.~(4): stellar mass; Col.~(5): ICL mass; Col.~(6): stellar and ICL mass to gas mass fraction.}
\begin{tabular}{c c c c c c}
\hline\hline
Cluster &   $L_{star}$  ($10^{14} L_{\odot}$)& $f_{s}$ & $M_{star}$ ($10^{14} M_{\odot}$)&  $M_{ICL}$& $M_{star+ICL}/M_{gas}$ \\
        & ($< r_{500}$)  & ($< r_{500}$) & ($< r_{500}$)&  ($< r_{500}$)& ($< r_{500}$) \\
\hline
A496  &  0.0233 $\pm$ 0.0021 & $0.20^{\ast}$ & 0.0351 $\pm$ 0.0066  & 0.0040 & 0.118  $\pm$ 0.020   \\
A1689 &  0.0599 $\pm$ 0.0023 & 0.15          & 0.0903 $\pm$ 0.020   & 0.010  & 0.060  $\pm$ 0.012   \\
A2050 &  0.0367 $\pm$ 0.0010 & 0.20          & 0.0553 $\pm$ 0.0015  & 0.062  & 0.108  $\pm$ 0.040  \\
A2631 &  0.0509 $\pm$ 0.0041 & 0.20          & 0.0768 $\pm$ 0.0062  & 0.0087 & 0.102  $\pm$ 0.022  \\
A2667 &  0.160  $\pm$ 0.046 & $0.20^{\ast}$  & 0.240  $\pm$ 0.073   & 0.027  & 0.237  $\pm$ 0.029 \\
\hline	
\end{tabular} 
 \\
\begin{minipage}{8.0truecm}
\footnotesize{$^\ast$ Not measured}\\
\end{minipage}
\label{StarMass}
\end{table*}

\section{Discussion}
\label{discus}

\begin{figure}[h!]
\centering
\includegraphics[width=0.40\textwidth,angle=90]{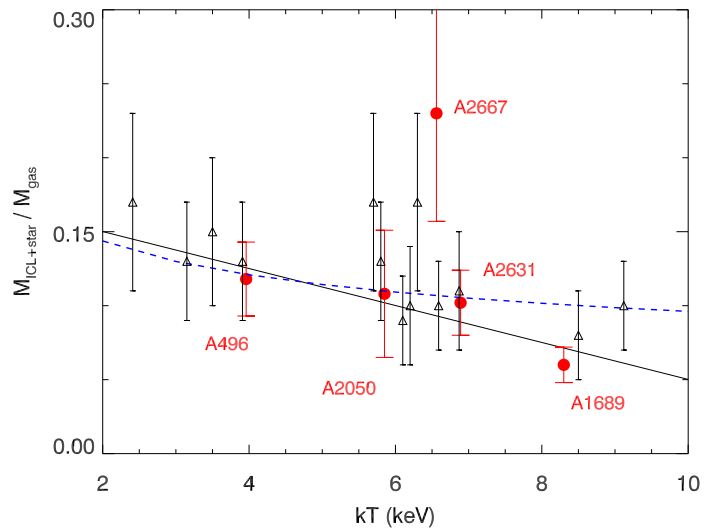}
\caption{ $M_{\rm ICL+star}/M_{\rm gas}$ as a function of temperature for A496, A1689, A2050, A2631 and A2667 (red points). Our best fit (Eq.~\ref{eq_ef}) is represented by the solid line. Black points were obtained from \citet{LMS} and the blue dot-dashed line was their best fit ($M_{\rm star}/M_{\rm gas} \propto kT^{0.23}$).} 
\label{EfkT}
\end{figure}

Combining X-ray and optical analysis we show in Fig.~\ref{EfkT} that for this sample the stellar-to-gas mass ratio is anti-correlated with X-ray temperature, as proposed by \citet{David90,LMS,Gonzales07}. \citet{Mohr99} found that clusters with temperature below 5 keV have a mean ICM mass fraction significantly lower than that of the hotter clusters, proposing an increase of the stellar mass in cooler clusters to maintain the baryon budget.

In Fig.~\ref{EfkT} we show the stellar-to-gas mass ratio as a function of the mean temperature of the cluster in our sample and for the clusters studied by \citet{LMS}. There is a clear decrease of $M_{\rm star}/M_{\rm gas}$ with increasing  gas temperature (which is indicative of the potential well), reflecting that the efficiency of converting gas into stellar matter is greater in cooler and smaller systems. 

The star formation efficiency in a galaxy can be defined in terms of the total stellar mass to gas mass ratio as stated above:
\begin{equation}
\epsilon=\bigg(\frac{M_{\rm ICL}+ M_{\rm star}}{M_{\rm star}+M_{\rm gas}+M_{\rm ICL}}\bigg).
\end{equation}

Although the number of clusters in this analysis is small, a Spearman test of correlation yields $\rho=-0.7$ with a confidence level of 81$\%$, with all data points included. Thus we propose a way to estimate the stellar content (diffuse stellar component plus the stars within galaxies) of a cluster based on its gas mass and temperature:
\begin{equation}
\frac{M_{\rm star}}{M_{\rm gas}}=0.18 - 0.012 \bigg(\frac{T_{\rm gas}}{\rm 1 keV}\bigg).
\label{eq_ef}
\end{equation} 

Differently from \citet{LMS}, we found a linear correlation between stellar-to-gas mass and temperature, which seems to well describe both samples.
 
The analysis of the stellar content of A2667
was not homogeneous with respect to the other four clusters, requiring several
reasonable assumptions. Its star formation efficiency appears in good
agreement with the relation defined by the remaining clusters at the
$2 \sigma$ confidence level.

We suggest one possible explanation for this trend in the context of the hierarchical scenario: during group mergers that form rich clusters, different gas galaxy stripping efficiencies are experienced. In more massive clusters, galaxies exhibit higher velocity dispersion and as a consequence gas stripping is stronger, diminishing star formation efficiency in galaxies.  Thus, physical mechanisms depending on virial mass, such as ram-pressure stripping, are driving galaxy evolution within clusters.

Only recently, due to advances in computing capabilities, cluster
 simulations have reached adequate levels of detail to trace star formation. Current numerical simulations use a quantity for the cosmic star formation rate that makes it possible to compare the mean efficiency of star formation between halos of different masses. Hydrodynamical simulations performed by \citet{Springel,Romeo05} appear to confirm that the stellar-to-gas mass ratio decreases with increasing temperature. However for \citet{Romeo05} this trend is driven by the fact that larger clusters better retain the ICM gas, so that the ICM to total mass ratio increases, while the stellar mass fraction remains fairly constant with cluster temperature. Although the latter behavior is in conflict with observations, the increase of the gas fraction with temperature was also found by \citet{AH07} who claim that a possible physical explanation for lower gas fraction in cooler systems is the higher amount of baryonic matter in the form of stars able to mantain the baryon budget constant.

\begin{figure}[t!]
\centering
\includegraphics[width=0.35\textwidth,angle=90]{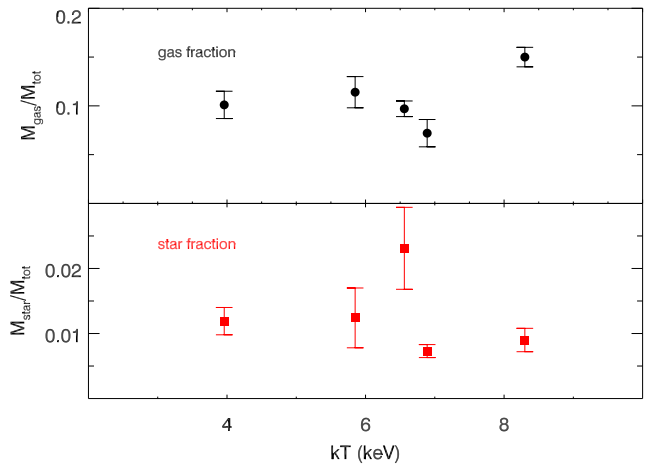}
\caption{Gas and total stellar fraction as a function of cluster temperature. The black points represent the gas fraction while the red ones are the stellar fraction. For the stellar fraction we used both stars in galaxies and the ICL contribution.}
\label{fbkt}
\end{figure}

The idea that the actual measurements of $f_{\rm gas}$ in clusters should be universal is a starting point for interesting cosmological tests. There has been some debate about the dependence of the gas fraction on temperature \citep{ME99,Mohr99,Roussel} and this discussion is still open. Our study suggests that the gas fraction is independent of the temperature, as shown in Fig.~\ref{fbkt} and pointed out in \citet{Allen04}. However, due to the limited sample we cannot draw a firm conclusion. To examine in a more systematic and uniform way the gas and the stellar fraction variation with temperature, a larger sample and stellar mass estimations are required.

\section{Conclusions}
\label{conc}

In this work we combined an X-ray with optical analysis to study the baryon content (stellar, intracluster stars and gas component) of five Abell clusters. Our results are robust because of homogenous data processing, a careful treatment of the stellar mass-to-light ratio of early and late-type galaxies and the use of XMM-Newton data to estimate the gas mass. We computed the stellar and the gas masses within the same radius.

From X-rays we derived the surface brightness and temperature profiles and we also computed the map of substructures for each cluster that gives hints about the dynamical state of each cluster. We show that the gas fraction obtained by means of the S\'{e}rsic equation is in agreement with values in the literature obtained with the double $\beta$-model. 

From optical data we derived the total luminosity and the late-type fraction to calculate stellar masses. We assumed a reasonable contribution of the ICL to the total luminosity.

We confirmed here, using a different technique, early findings by \citet{David90,LMS,Gonzales07} regarding the dependence of the stellar-to-gas mass ratio on the temperature. \citet{LMS} have quantified this ratio which agrees with our results. 

 We showed here that, for this sample of clusters, the efficiency of galaxy formation can vary from 6$\%$ to 14$\%$ proving that the star formation efficiency depends on the environment, as hydrodynamical simulation performed by \citet{Springel} have confirmed. 

\begin{acknowledgements}
The authors acknowledge financial support from the Brazilian agencies FAPESP (grants: 03/10345-3, 01/07342-7) and CAPES (grant: BEX1468/05-7) and the Brazilian-French collaboration CAPES/Cofecub (444/04). We thank the anonymous referee who carefully read this paper and made very useful comments. We also thank Gwenael Bou\'{e} for making the A496 photometric catalog available before publication, Renato Dupke, Walter Santos, Florence Durret, Cristophe Adami and Graziela Keller for constructive suggestions and discussion. We also thank Brent Groves and Dave Stock for proof reading this manuscript.
Based on observations obtained with MegaPrime/MegaCam, a joint project of CFHT and CEA/DAPNIA, at the Canada-France-Hawaii Telescope (CFHT) which is operated by the National Research Council (NRC) of Canada, the Institute National des Sciences de l'Univers of the Centre National de la Recherche Scientifique of France, and the University of Hawaii.
\end{acknowledgements}

\bibliographystyle{aa}
\bibliography{9168}

\end{document}